\documentclass[twocolumn,pre,aps,showpacs,floatfix,superscriptaddress]{revtex4}
\usepackage{epsfig}
\usepackage{graphicx}
\begin{document}

\widetext
\title{Maximal width of the separatrix chaotic layer}

\author{S.M. Soskin}
%\altaffiliation{Also at Physics Department, Lancaster University, UK}
\affiliation{Institute of Semiconductor Physics, National
Academy of Sciences of Ukraine, 03028 Kiev, Ukraine}
\affiliation{Abdus Salam ICTP, 34100 Trieste, Italy}
\affiliation{Physics Department, Lancaster University, Lancaster LA1 4YB, UK}
\author{R. Mannella}
\affiliation{Dipartimento di Fisica, Universit\`{a} di Pisa, 56127
Pisa, Italy}

\begin{abstract}
The main goal of the paper is to find the {\it absolute maximum} of the width
of the separatrix chaotic
layer as function of the frequency of the
time-periodic perturbation of a one-dimensional Hamiltonian system possessing a separatrix, which is one of the major unsolved problems in the theory of separatrix chaos.
For a given small amplitude of the perturbation, the width is shown to possess
sharp peaks in the range from logarithmically small to moderate
frequencies. These peaks are universal, being the consequence of the
involvement of the nonlinear resonance dynamics into the separatrix
chaotic motion.
Developing further the approach introduced in the
recent
paper by Soskin et al. ({\it PRE} {\bf 77}, 036221 (2008)),
we derive leading-order asymptotic expressions for the shape of the
low-frequency peaks. The maxima of the peaks, including in particular the {\it absolute maximum} of the width, are proportional to the
perturbation amplitude times either a logarithmically large factor
or a numerical, still typically large, factor, depending on the type of system.
Thus, our theory predicts that the maximal width of the chaotic layer may be much larger than that predicted by former theories.
The theory is verified in simulations.
An application to the facilitation of global chaos onset is discussed.
\end{abstract}

 \maketitle

\section{Introduction}

Separatrix chaotic layers (SCLs) play a fundamental role for Hamiltonian
chaos and may be important in a broad variety of subjects in
physics and astronomy
\cite{ZF:1968,Chirikov:79,lichtenberg_lieberman,Zaslavsky:1991,zaslavsky:1998,zaslavsky:2005,abdullaev,gelfreich,treschev,nine_prime,nine_prime_prime}.
One of the most important characteristics of the layer is its
width in energy or in related quantities.
It
can be easily found {\it numerically} by
means of
integration of the Hamiltonian
equations with a set of initial conditions in the vicinity of the
separatrix.
But it is important also to be able to
%describe various properties of the SCL
find it
{\it theoretically}. There is a long and rich history
of the corresponding studies. The results may be classified as follows.

{\bf 1. Heuristic analytic results.}

\noindent
Consider a 1D Hamiltonian system perturbed by a weak time-periodic
perturbation:

\begin{eqnarray}
&& H=H_0(p,q)+hV(p,q,t),\\
&& V(p,q,t+2\pi/\omega_f)=V(p,q,t),\quad\quad h\ll 1,
\nonumber
\end{eqnarray}

\noindent where
%the unperturbed Hamiltonian
$H_0(p,q)$ possesses a
separatrix and, for the sake of notation compactness, all relevant
parameters of $H_0$ and $V$, except possibly $\omega_f$, are
assumed to be $\sim 1$.

There were a few heuristic criteria set by physicists (see e.g.
\cite{ZF:1968,Chirikov:79,lichtenberg_lieberman,Zaslavsky:1991,zaslavsky:1998,zaslavsky:2005})
%[1-3])
which
gave qualitatively similar results for the SCL width $\Delta E$ in terms of
energy $E\equiv H_0(p,q)$:

\begin{eqnarray}
&& \Delta E\equiv\Delta E(\omega_f)\sim  \omega_f\delta,\quad\quad
\\
&& \delta \equiv h|\epsilon|,
\nonumber
\\
&&
%\quad\quad
|\epsilon|\stackrel{<}{\sim} 1\quad\quad\quad\quad\quad\quad\quad\quad\quad \quad\quad\quad\quad  {\rm for}\quad
\omega_f\stackrel{<}{\sim} 1,
\nonumber
\\
&&
|\epsilon|\propto \exp(-a\omega_f)\ll 1 \quad \quad (a\sim 1) \quad \quad {\rm for}\quad
\omega_f\gg 1.
\nonumber
\end{eqnarray}

\noindent The quantity $\delta \equiv h|\epsilon|$ is called the {\it separatrix
split} \cite{zaslavsky:1998}
(see also Eq. (4) below):
%below):
it determines the maximum distance between the perturbed incoming and
outgoing separatrices
\cite{ZF:1968,Chirikov:79,lichtenberg_lieberman,Zaslavsky:1991,zaslavsky:1998,zaslavsky:2005,abdullaev,gelfreich,treschev}.

It follows from (2) that the maximum of $\Delta E$
lies in the frequency range $\omega_f\sim 1$ while the maximum itself is $\sim
h$:

\begin{equation}
\Delta E_{\max}\equiv\max_{\omega_f}\{\Delta E(\omega_f)\}\sim h,
%\quad\quad\quad
\quad\quad\quad \omega_f^{(\max)}\sim 1.
\end{equation}

{\bf 2. Mathematical and accurate physical results.}

\noindent
Many papers studied the SCL by mathematical and accurate physical methods. %typical of mathematicians.

For the range $\omega_f\gg 1$, there were many works
studying the separatrix splitting (see the review \cite{gelfreich} and references therein)
and the SCL width in terms of the normal coordinates
(see the review \cite{treschev} and references therein). Though
quantities studied in these works typically differ from those
studied by physicists
\cite{ZF:1968,Chirikov:79,lichtenberg_lieberman,Zaslavsky:1991,zaslavsky:1998,zaslavsky:2005},
they
implicitly confirm the main qualitative conclusion from the heuristic formula (2)
in the high frequency range: if $\omega_f\gg 1$
the SCL width is exponentially small.

There were also several works studying the SCL in the opposite
(i.e. adiabatic) limit $\omega_f\rightarrow 0$: see e.g.
\cite{Neishtadt:1986,E&E:1991,Neishtadt:1997,prl2005,13_prime}
and
references therein. In the context of the SCL width, it is most
important that $\Delta E(\omega_f\rightarrow 0)\sim h$ for most of the systems
%[6-8].
\cite{Neishtadt:1986,E&E:1991,Neishtadt:1997}.
For a particular
class of systems, namely for ac-driven spatially periodic systems
(e.g. the ac-driven pendulum), the width of the SCL part above the
separatrix diverges in the adiabatic limit \cite{prl2005,13_prime}: the
divergence develops for $\omega_f\ll 1/\ln(1/h)$.

Finally, there is a
qualitative estimation
%study
of the SCL width for the range $\omega_f\sim 1$
%by one of the mathematical methods
within the Kolmogorov-Arnold-Moser (KAM) theory
\cite{treschev}
while
the
quantitative estimation within the KAM theory appears to be very difficult for this frequency range \cite{vasya}.
It follows from the results in
\cite{treschev} that the width in this range is of the order of the
separatratrix split while the latter is of the order of $h$.

Thus, from the above results, it could seem to follow that, for all systems except the ac-driven spatially periodic systems,
the maximum of the SCL width is $\sim h$ and occurs in the range
$\omega_f\sim 1$, quite in agreement with the heuristic result
(3). Even for the ac-driven spatially periodic
systems, this conclusion could seem to apply to the width of the SCL
part below the separatrix, for the whole frequency range, and to the width of the SCL part above
the separatrix, for $\omega_f \stackrel{>}{\sim} 1/\ln(1/h)$.

%\noindent
{\bf 3. Numerical evidences of high peaks in
$\Delta E(\omega_f)$
and their rough estimates.}

\noindent
The
above
conclusion does not agree with several
numerical studies carried out during the last decade (see e.g.
\cite{prl2005,13_prime,shevchenko:1998,luo1,soskin2000,luo2,vecheslavov,shevchenko})
%[9-15])
which have revealed the existence of sharp
peaks in $\Delta E(\omega_f)$
in the frequency range
$1/\ln(1/h)\stackrel{<}{\sim}\omega_f\stackrel{<}{\sim} 1$
the heights of which greatly exceed $h$
(see also Figs. 2, 3, 5, 6 below).
Thus, the peaks represent the general {\it dominant feature} of
the function $\Delta E(\omega_f)$.
%Intuitively, the
The
peaks were related by the authors of
%[10-15]
\cite{shevchenko:1998,luo1,soskin2000,luo2,vecheslavov,shevchenko}
to the absorption of nonlinear resonances by the SCL. For some partial
case,
rough analytic estimates for the position and magnitude of the
peaks were made in
\cite{shevchenko:1998,shevchenko}.

%\noindent
{\bf 4. Approach to an accurate description of the peaks.}

\noindent
Accurate analytic estimates for the peaks were lacking.
It is explicitly stated in the review \cite{luo2} that the search for the mechanism
of
the involvement of resonances
into the separatrix chaos and an accurate analytic
description of the peaks
are being among the most important and challenging tasks in the
study of separatrix chaos.
The first step
towards this accomplishment
was done in the recent papers \cite{pre2008,proceedings}, where a new approach to the
theoretical treatment of the separatrix chaos for the relevant frequency range was
developed and applied to the problem of the onset of global chaos between two close separatrices.
An application of the approach to the, more common, single-separatrix cases was only discussed in \cite{pre2008,proceedings}.

\vspace*{0.2cm}

The {\it present} paper formulates the basic ideas of the approach in terms more general than \cite{pre2008,proceedings} and, on the basis of this approach, develops the {\it first ever} accurate theoretical description of the peaks i.e. of the SCL width as a function of frequency in the range of the {\it maximum} of the width, which is the most important range from the physical point of view.
In particular, we show that the maximal width of the separatrix chaotic layer may be much larger than it was assumed before. In the latter context, all systems are classified by us into two different types: for systems of type I, the ratio between the maximal width and the perturbation amplitude $h$ {\it logarithmically diverges} in the asymptotic limit $h\rightarrow 0$ while, for systems of type II, it asymptotically approaches a {\it constant} (still large, typically).

Though the form of our treatment differs from typical forms of mathematical theorems in this subject (cf. \cite{gelfreich,treschev}), the results yield the {\it exact} leading-order term in the asymptotic expansion of the width in the parameter of smallness $\alpha\equiv1/\ln(1/h)$. Our theory is in excellent agreement with the results of numerical integration of the equations of motion.

Sec. II describes the basic ideas of the approach.
Sec. III presents the classification into two types of systems,
using rough estimates. Sec. IV develops the leading-order
asymptotic theory for an archetypal example of type I and
compares it with the numerical integration of Hamiltonian equations of motion. Sec. V develops
the leading-order asymptotic theory for two archetypal examples of
type II and compares it with the numerical integration. Next-order corrections are estimated in Sec. VI. Discussion of a few other issues, including in particular an application to the global chaos onset, is presented in
Sec. VII. Conclusions are drawn in Sec. VIII.

\section{Basic ideas of the approach}

The new approach, which is developed in \cite {pre2008,proceedings} and here, may be briefly formulated as the matching between the discrete chaotic dynamics of the separatrix map in the immediate vicinity of the separatrix and the continuous regular-like dynamics of the resonance Hamiltonian beyond the close vicinity of the separatrix. The present section describes the general features of the approach in more details.

The motion near the separatrix may be approximated by the
{\it separatrix map} (SM)
\cite{ZF:1968,Chirikov:79,lichtenberg_lieberman,Zaslavsky:1991,zaslavsky:1998,zaslavsky:2005,abdullaev,treschev,shevchenko:1998,shevchenko,pre2008,proceedings,vered}.
%[1-3,5,16,17].
It was introduced for the first time in \cite{ZF:1968} and its various modifications were used in many studies afterwards, sometimes being called as the {\it whisker map}. It was re-derived in \cite {vered} rigorously, as the leading-order approximation of the motion near the separatrix in the asymptotic limit $h\rightarrow 0$, and an estimate of the errors was carried out too (see also \cite{treschev} and references therein).

We remind the main ideas which allow one to introduce the SM
\cite{ZF:1968,Chirikov:79,lichtenberg_lieberman,Zaslavsky:1991,zaslavsky:1998,zaslavsky:2005,abdullaev,treschev,pre2008,proceedings,vered}. 
For the sake of simplicity, let us consider a perturbation $V$ that does not depend on the momentum: $V\equiv V(q,t)$.
A system with an energy close to the separatrix value
%$E_s$
spends most of the time in the vicinity of the saddle(s), where the velocity is
exponentially small. 
Differentiating $E\equiv H_0(p,q)$ with respect to time and allowing for the equations of motion of the system (1), we can show that $\dot{E}=\dot{q}\partial V/\partial q\propto \dot{q}$.
Thus, the perturbation can significantly change the energy
only when the velocity is not small i.e. during the relatively short intervals while
the system is away from the saddle(s): these intervals correspond to {\it pulses}
of velocity as a function of time. Consequently, it is possible to approximate the
continuous Hamiltonian dynamics by a discrete dynamics which maps the energy $E$,
the perturbation angle $\varphi\equiv \omega_f t$ and the velocity sign
$\sigma\equiv{\rm sgn}(\dot{q})$ from pulse to pulse.

The
actual form of the SM may
vary, depending on the system under study, but its features, relevant in
the present context, are similar for all systems. For the
sake of clarity, let us consider the explicit case when
the separatrix of $H_0(p,q)$ possesses
%of the shape \lq\lq eight'' i.e.
a single saddle and two symmetric loops while $V=q\cos(\omega_ft)$.
%(like in a double-well potential system).
Then the SM reads \cite{pre2008}:
%(cf. %\cite{ZF:1968,Chirikov:79,lichtenberg_lieberman,Zaslavsky:1991,zaslavsky:1998,zaslavsky:2005,abdullaev,treschev,vered}):

\begin{eqnarray}
&&E_{i+1}=E_i+\sigma_ih\epsilon\sin(\varphi_i),
\\
&&\varphi_{i+1}=\varphi_i+\frac{\omega_f\pi(3- {\rm
sgn}(E_{i+1}-E_s)) }{2\omega(E_{i+1})}, \nonumber
\\
&&\sigma_{i+1}=\sigma_i \, {\rm sgn}(E_s-E_{i+1}), \quad\quad |\sigma_i|=1, \nonumber
\\
&&\quad\quad\quad\epsilon \equiv \epsilon(\omega_f)=
\nonumber
\\
&&\quad\quad\quad
{\rm sgn}\left(\left.\frac{\partial
H_0}{\partial p}\right|_{t\rightarrow -\infty}\right)
%\sigma_i
\int_{-\infty}^{\infty}
%_{i{\rm th}\quad{\rm pulse}}
{\rm d}t\;\left.\frac{\partial H_0}{\partial p}\right|_{E_s}\sin(\omega_ft)
,
\nonumber
\\
&&\quad\quad\quad E_i\equiv \left.H_0(p,q)\right|_{t_i-\Delta},
\nonumber
\\
&&\quad\quad\quad\varphi_i\equiv
\omega_ft_i,
\nonumber
\\
&&\quad\quad\quad\sigma_i\equiv{\rm sgn}\left(\left.\frac{\partial
H_0}{\partial p}\right|_{t_i}\right), \nonumber
\end{eqnarray}

\noindent where $E_s$ is the separatrix energy, $\omega(E)$
is the frequency of oscillation with energy $E$ in the
unperturbed case (i.e. for $h=0$), $t_i$ is the instant corresponding to the $i$-th turning point in the trajectory $q(t)$, and $\Delta$ is an arbitrary value from the range of time intervals which greatly exceed the characteristic duration of the velocity pulse while being much smaller than the interval between the subsequent pulses
\cite{ZF:1968,Chirikov:79,lichtenberg_lieberman,Zaslavsky:1991,zaslavsky:1998,zaslavsky:2005,abdullaev,treschev,vered}.

Consider the two most general ideas of our approach.

{\bf 1. If a trajectory of the SM includes a state with $E=E_s$ and an arbitrary $\varphi$ and $\sigma$, then this trajectory is
%{\it chaotic}.}
chaotic.}
Indeed, the angle $\varphi$ of such a state is not correlated with
the angle of the state at the previous step of the map, due to the
divergence of $\omega^{-1}(E\rightarrow E_s)$.
Therefore, the angle at the previous step may deviate from a multiple of $2\pi$ by an arbitrary value and,
hence, the energy of the state at the previous step
may deviate from $E_s$ by an arbitrary value within the interval
$[-h|\epsilon|,h|\epsilon|]$. The
%variable
velocity sign
$\sigma$ is not correlated with that at
the previous step either \cite{22_prime}.
Given that a regular trajectory of the SM cannot include a step where all three variables of the SM change random-like, we conclude that such a trajectory is chaotic.

Though the above arguments appear to be obvious, they may not be considered as a
mathematically rigorous proof, so that the statement about the chaotic nature of the
SM trajectory which includes any state with $E=E_s$ should be considered as a
conjecture supported by the above arguments and by the results of the numerical
iteration of the SM. Possibly, the mathematically rigorous proof should involve an
analysis of the Lyapunov exponents for the SM (cf. \cite{lichtenberg_lieberman}) but
this appears to be a technically difficult problem. We emphasize however that the
rigorous proof of the conjecture is not crucial for the validity of the main results
of the present paper, namely of the {\it leading} terms in the asymptotic expressions
describing the peaks of the SCL width as a function of the perturbation frequency.
It will be obvious from the next item
that in order
to derive the leading term it is sufficient to know that the chaotic trajectory does visit areas of the phase space where the energy deviates from the separatrix by values of the order of the separatrix split $\delta\equiv h|\epsilon|$, which is a widely accepted fact
\cite{ZF:1968,Chirikov:79,lichtenberg_lieberman,Zaslavsky:1991,zaslavsky:1998,zaslavsky:2005,abdullaev,gelfreich,treschev}.

{\bf 2.} As well known
%[1-3,10,15-18],
\cite{ZF:1968,Chirikov:79,lichtenberg_lieberman,Zaslavsky:1991,zaslavsky:1998,zaslavsky:2005,abdullaev,gelfreich,treschev,shevchenko:1998,shevchenko,pre2008,proceedings}, at the leading-order approximation
the frequency of eigenoscillation as function of the energy near the
separatrix
is proportional to the reciprocal of the logarithmic factor

%\begin{eqnarray}
%&&\omega(E)=\frac{b\pi\omega_0}{\ln\left(\Delta H/|E-E_s|\right)},
%\quad \quad b=\frac{3- {\rm sign}(E-E_s)}{2},
%\\
%&& |E-E_s|\ll\Delta H\sim E_s-E_{st}^{(1)}\sim
%E_s-E_{st}^{(2)}\nonumber
%\end{eqnarray}

\begin{eqnarray}
&&\omega(E)=\frac{b\pi\omega_0}{\ln\left(\frac{\Delta H}{|E-E_s|}\right)},
%\quad
\quad  b=\frac{3- {\rm sgn}(E-E_s)}{2},
\\
%\quad\quad
&&|E-E_s|\ll\Delta H\equiv E_s-E_{st},
\nonumber
\end{eqnarray}

\noindent where $E_{st}$ is the energy of the stable states.

Given that the argument of the logarithm is large in the relevant
range of $E$, the function $\omega (E)$ is nearly constant for a
%rather significant
substantial
variation of the argument. Therefore, {\bf as the SM
maps the state $(E_0=E_s,\varphi_0,\sigma_0)$ onto the state with
$E=E_1\equiv E_s+\sigma_0 h\epsilon\sin(\varphi_0)$, the value of
$\omega(E)$ for the given ${\rm
sgn}(\sigma_0\epsilon\sin(\varphi_0))$ is nearly the same for most
of the angles $\varphi_0$ (except in the close vicinity of multiples of $\pi$)},
namely

\begin{eqnarray}
&&\omega(E)\approx\omega_r^{(\pm)},
\\
&&\omega_r^{(\pm)}
\equiv\omega(E_s\pm
h),
%\epsilon(\omega_f=\omega(E_s+h))
%\quad\quad {\rm for}
\quad\quad {\rm
sgn}(\sigma_0\epsilon\sin(\varphi_0))=\pm 1.
\nonumber
\end{eqnarray}

Moreover, if the deviation of the SM trajectory from the
separatrix increases further, $\omega(E)$ remains close to
$\omega_r^{(\pm)}$ provided the deviation is not too large,
namely if $\ln(|E-E_s|/h)\ll\ln(\Delta H/h)$. If
$\omega_f\stackrel{<}{\sim}\omega_r^{(\pm)}$, then the evolution of
the SM (4) may be regular-like for a long time until the energy
returns
%back
to the close vicinity of the separatrix, where the
%correlation of the
trajectory is
chaotized.
%again lost.
Such a behavior is especially
pronounced if the perturbation frequency is close to
$\omega_r^{(+)}$ or $\omega_r^{(-)}$ or to one of their multiples of
relatively low order:
% \cite{24_prime}:
the resonance between the perturbation and the
eigenoscillation gives rise to an accumulation of energy changes
for many steps of the SM, which results
%on
in
a deviation of $E$ from
$E_s$
%which
that
greatly exceeds the separatrix split $h|\epsilon|$.
Consider a state at the boundary of the SCL.
The deviation of energy of such a state from $E_s$ depends on its position at the boundary. In turn, the maximum deviation is a function of $\omega_f$. The latter function possesses the absolute maximum at $\omega_f$
close to
%$\omega_r^{(+/-)}$ ,
$\omega_r^{(+)}$ or $\omega_r^{(-)}$ typically \cite{24_prime},
for the
%upper/lower boundary of the SCL.
upper or lower boundary of the SCL respectively.
This corresponds
to the absorption of the, respectively upper and lower, 1st-order nonlinear resonance by the SCL.

The above intuitive idea has been explicitly confirmed in \cite{pre2008}: it has been shown in the Appendix of \cite{pre2008} that, in the relevant range of energies, the separatrix map can be reduced to the system of two differential equations which are indentical to the equations of motion of the auxiliary resonance Hamiltonian which describes the resonance dynamics in terms of the conventional canonically conjugate slow variables, action $I$ and slow angle $\tilde{\psi}\equiv n \psi-\omega_ft$ where $\psi$ is the angle variable
\cite{Chirikov:79,lichtenberg_lieberman,Zaslavsky:1991,zaslavsky:1998,zaslavsky:2005,abdullaev} (see Eq. (16) below) while $n$ is the relevant resonance number i.e. the integer number closest to the ratio $\omega_f/\omega_r^{(\pm)}$.

\vspace*{0.5cm}

Thus, the result of the matching between the discrete chaotic dynamics of the SM and the continuous regular-like dynamics of the resonance Hamiltonian is the following \cite{pre2008}.
After the chaotic trajectory of the SM visits
any state on the separatrix, the system transits in one step of the
SM to a given upper or lower curve in the $I-\tilde{\psi}$ plane which has been labelled \cite{pre2008} respectively upper or lower {\it generalized separatrix
split} (GSS) curve \cite{22_prime_prime}:

\begin{equation}
E=E_{GSS}^{(\pm)}(\tilde{\psi})\equiv
E_s\pm\delta|\sin(\tilde{\psi})|,\quad\quad \delta\equiv
h|\epsilon|,
\end{equation}

\noindent where $\delta$ is the conventional separatrix split
\cite{zaslavsky:1998} while $\epsilon$ is the amplitude of the Melnikov-like integral defined in Eq. (4) above (cf.
\cite{ZF:1968,Chirikov:79,lichtenberg_lieberman,Zaslavsky:1991,zaslavsky:1998,zaslavsky:2005,abdullaev,gelfreich,treschev,shevchenko:1998,vecheslavov,shevchenko,pre2008,proceedings}),
and
the angle $\tilde{\psi}$ may take any value from one of the two ranges: either $[0,\pi]$ or $[\pi, 2\pi]$ \cite{footnote_new}.

After that, because of the closeness of $\omega_f$ to the $n$-th harmonic
of $\omega(E)$ in the relevant range of $E$ \cite{footnote_new_former},
for a relatively long time the system follows
the {\it nonlinear resonance} (NR) dynamics (see Eq. (16) below),
during which the deviation of the energy from the separatrix value grows,
greatly exceeding $\delta$ for most of the trajectory. As time goes on,
$\tilde{\psi}$ is moving and, at some point, the
deviation in energy from the separatrix value begins to decrease.
This decrease lasts until the
system hits the GSS curve,
%(in the interval neighbouring $\pi$),
after which it returns to the separatrix just for one step of the
separatrix map. At the separatrix, the slow angle $\tilde{\psi}$ is
chaotized, so that a new stage of evolution similar to the one just
described occurs, i.e. the nonlinear resonance dynamics starting from the GSS curve with a new (random) value of 
%the \lq\lq initial'' value of 
$\tilde{\psi}$.

Of course, the SM cannot describe the
%change
variation
of the energy during the velocity
pulses (i.e. in between instants relevant to the SM): in some cases this
%change
variation
can be  comparable with the change within the SM dynamics. This additional
%change
variation
will be taken into account below, where relevant (see Sec. V below).

One might argue that, even for the instants relevant to the SM, the SM describes the original Hamiltonian dynamics only approximately \cite{vered} and may therefore miss some fine details of the motion: for example, the above picture does not include small windows of stability on the very separatrix. However these fine details are irrelevant in the present context, in particular the relative portion of the windows of stability on the separatrix apparently vanishes in the asymptotic limit $h\rightarrow 0$.

The boundary of the SM chaotic layer is formed by those parts of the SM chaotic trajectory which deviate from the separatrix more than others.
As follows from the structure of the chaotic trajectory described
above, the upper/lower boundary of the SM chaotic layer is formed in one of the two following ways
\cite{pre2008,proceedings}: 1) if there exists a {\it
self-intersecting} resonance trajectory (in other words, the
resonance separatrix) the lower/upper part of which
(i.e. the part situated below/above the self-intersection) touches or
intersects the upper/lower GSS curve while the upper/lower part does not,
then the upper/lower boundary of the layer is formed by the
upper/lower part of this self-intersecting trajectory (Figs. 1(a) and 1(b));
2) otherwise the boundary is formed by the resonance trajectory {\it tangent} to
the GSS curve (Fig. 1(c)). It is shown
%in the next sections
below
that, in both cases,
the variation of the energy along the resonance trajectory is larger
than the separatrix split $\delta$ by a logarithmically large factor
$\propto \ln(1/h)$. Therefore, over the boundary of the SM chaotic layer
%in the asymptotic limit $h\rightarrow 0$,
%the largest
%maximum
the largest
%(over the boundary of the SM chaotic layer)
deviation of the energy from the separatrix value, $\Delta E^{(\pm)}_{sm}$, may be
taken, in the leading-order approximation, to be equal to the
%maximal
largest
variation of the energy along the resonance trajectory
forming the boundary, while the latter trajectory can be entirely described within the resonance Hamiltonian formalism.

Finally, we mention in this section that, as obvious from the above description of the
boundary, $\Delta E^{(\pm)}_{sm}\equiv \Delta
E^{(\pm)}_{sm}(\omega_f)$ possesses a local maximum $\Delta
E^{(\pm)}_{\max,sm}$ at $\omega_f$ for which the resonance
separatrix just {\it touches} the corresponding GSS curve (see Fig.
1(a)).

\begin{figure}[tb]
\includegraphics*[width = 6. cm]{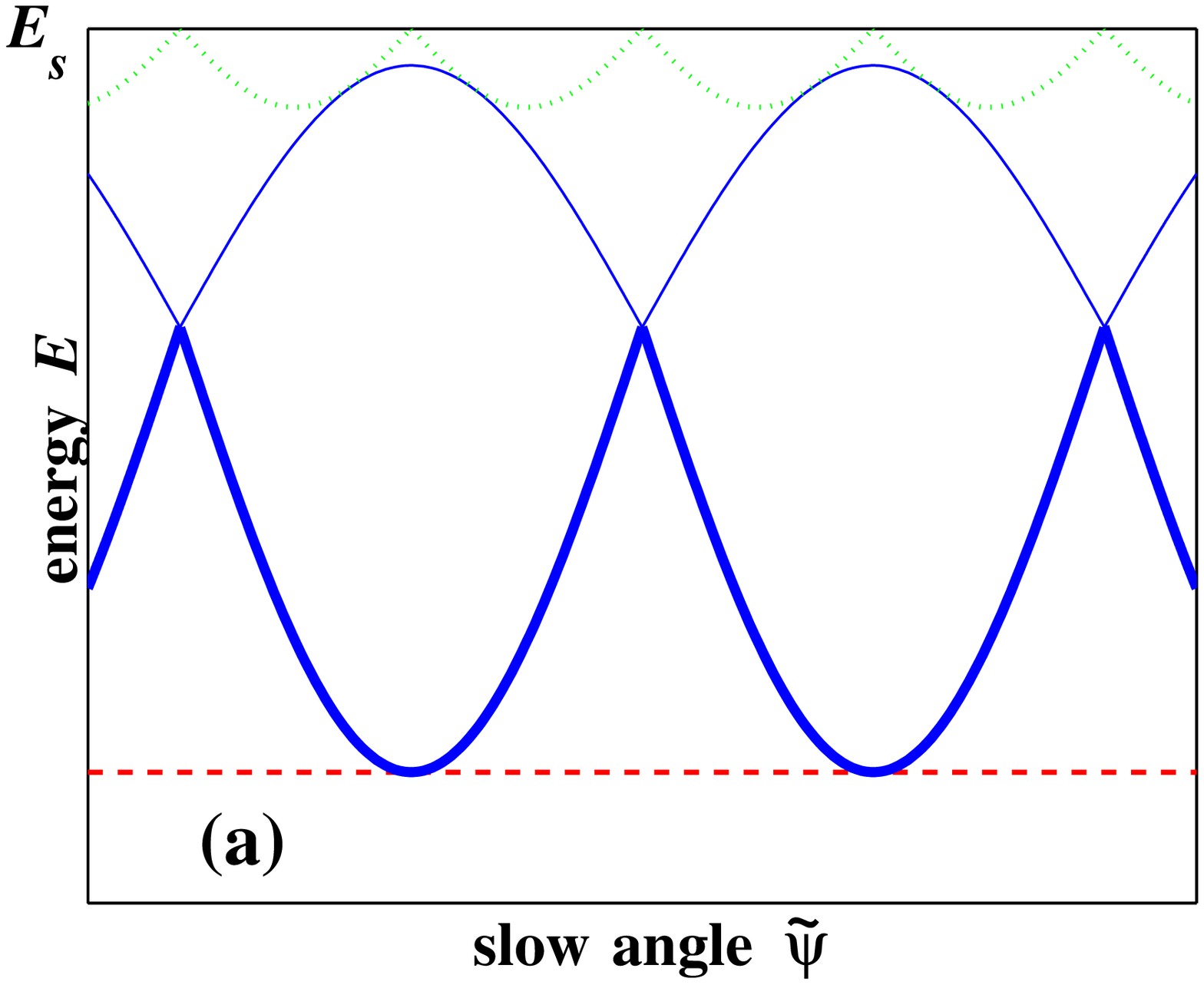}
\vskip 0.2 cm
\includegraphics*[width = 6. cm]{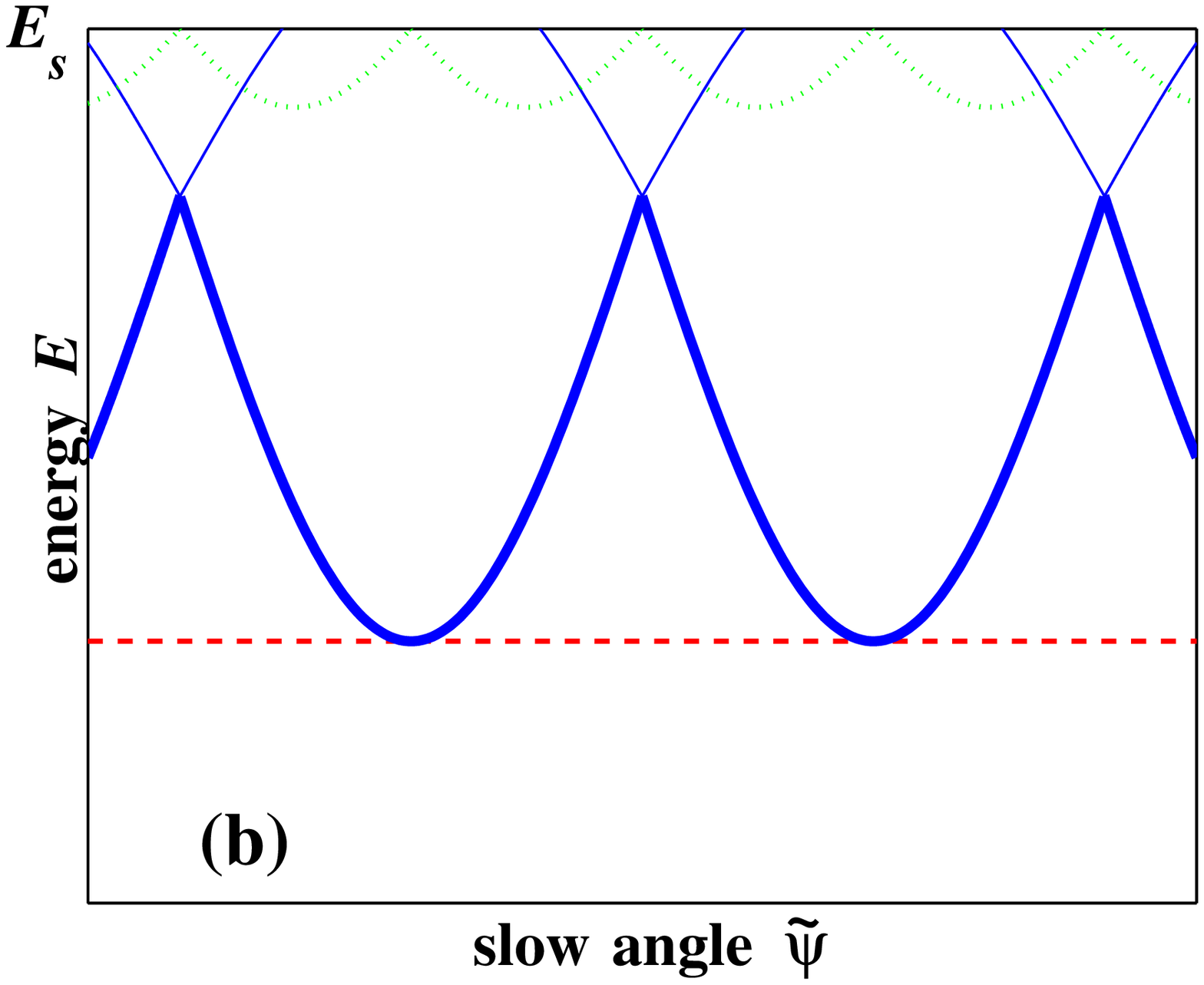}
\vskip 0.2 cm
\includegraphics*[width = 6. cm]{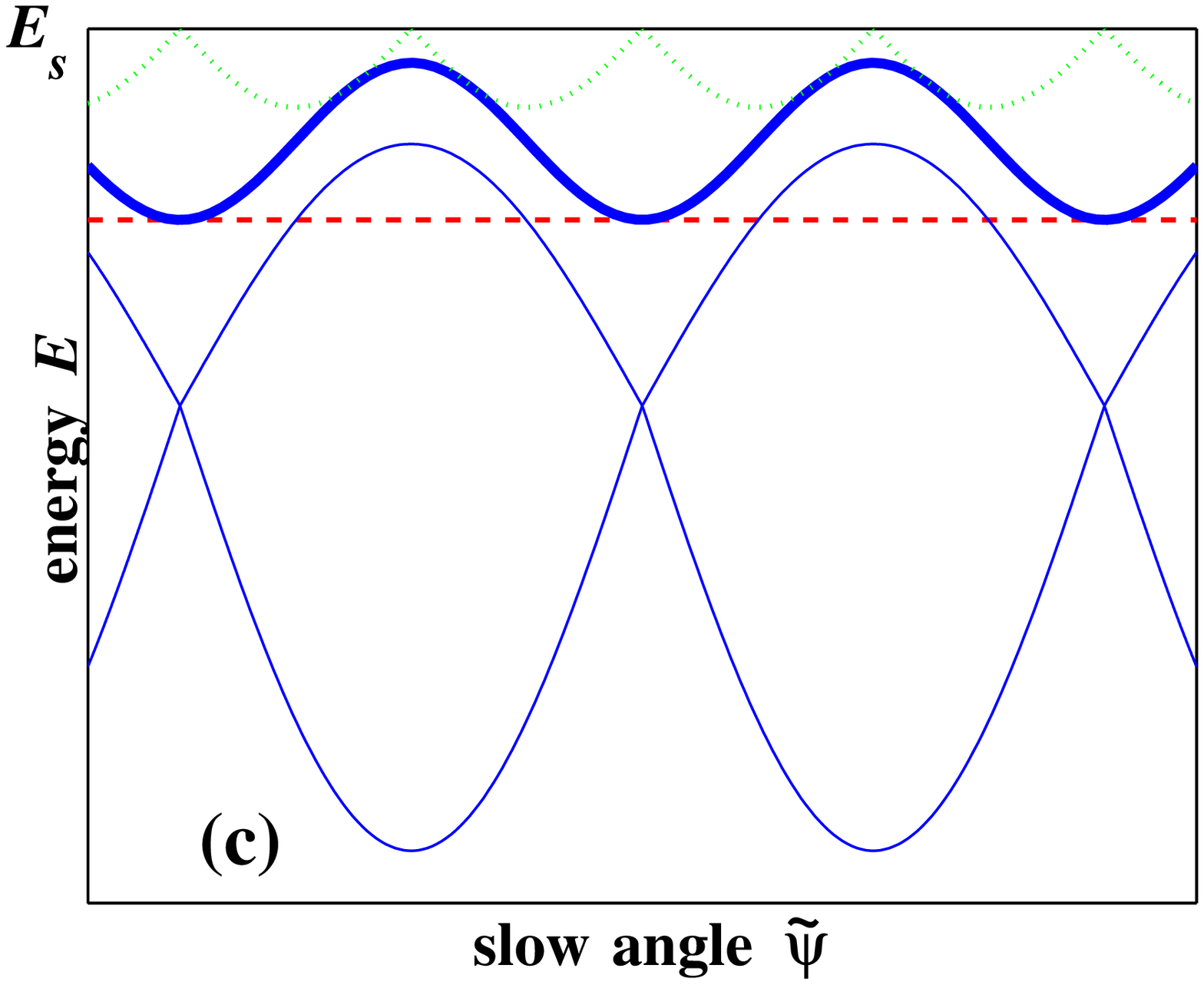}
\vskip 0.2 cm \caption {A schematic figure illustrating the
formation of the peak of the
 function $\Delta E^{(-)}_{sm}(\omega_f)$: (a) $\omega_f=\omega_{\max}$;
 (b) $\omega_f<\omega_{\max}$; (c) $\omega_f>\omega_{\max}$.
 The relevant (lower) GSS curve is shown by the dotted line. The relevant trajectories of the resonance Hamiltonian
 are shown by solid lines. The lower boundary of the layer
 is marked by a thick solid line: in (a) and (b)
 the lower boundary is formed by the lower part of the resonance separatrix while,
 in (c) it is formed by the resonance trajectory tangent to the GSS curve.
 The dashed line marks, for a given
 $\omega_f$, the maximal
deviation of the lower boundary from the separatrix energy $E_s$. }
\label{aba:fig1}
\end{figure}

\section{Rough estimates. Classification of systems.}

As obvious from Sec. II above,
$\Delta E^{(\pm)}_{\max,sm}$
is equal, in the leading order, to the width $\Delta E_{NR}$ of the nonlinear
resonance which touches the separatrix.
Let us make a rough estimate of $\Delta E_{NR}$: it will turn out
that it is possible to classify all systems into two different types.
With this aim, we expand the perturbation $V$ into a Fourier
series in $t$ and a Fourier series in $\psi$:

\begin{eqnarray}
V&&\equiv  \frac{1}{2}\sum_l V^{(l)}(E,\psi)\exp(-{\rm
i}l\omega_ft)+{\rm c.c.}\nonumber\\&&\equiv\frac{1}{2}\sum_{l,k}
V^{(l)}_{k}(E)\exp({\rm i}(k\psi-l\omega_ft))+{\rm c.c.}
\end{eqnarray}

As in the standard theory of a nonlinear resonance
\cite{Chirikov:79,lichtenberg_lieberman,Zaslavsky:1991,zaslavsky:1998,zaslavsky:2005},
%(its mathematical rigour is based on the averaging method \cite{bogmit}),
let us single out the relevant $V^{(L)}_{K}$ for a given peak, and
denote its absolute value by $V_0$:

\begin{equation}
V_0(E)\equiv |V^{(L)}_{K}(E)|.
\end{equation}

Let us now roughly estimate the width of the resonance, using the
pendulum approximation of the resonance dynamics
\cite{Chirikov:79,lichtenberg_lieberman,Zaslavsky:1991,zaslavsky:1998,zaslavsky:2005,abdullaev}:

\begin{equation}
\Delta E_{NR}\sim \sqrt{\frac{8hV_0\omega_f}{|{\rm d}\omega/{\rm
d}E|}}.
\end{equation}

Of course, this approximation assumes the constancy of ${\rm
d}\omega/{\rm d}E$ in the resonance range of energies, while it is
not so in our case: $\omega(E)\propto 1/\ln(1/|E-E_s|)$ in the
vicinity of the separatrix
\cite{ZF:1968,Chirikov:79,lichtenberg_lieberman,Zaslavsky:1991,zaslavsky:1998,zaslavsky:2005,abdullaev,treschev,shevchenko:1998,vecheslavov,shevchenko,pre2008,proceedings},
so that the relevant derivative $|{\rm d}\omega/{\rm d}E|\sim
(\omega_r^{(\pm)})^2/(\omega_0|E-E_s|)$ strongly varies within the resonance
range. However, for our rough estimate we may use the maximal value
of $|E-E_s|$, which is equal to $\Delta E_{NR}$ approximately. If
$\omega_f$ is of the order of $\omega_r^{(\pm)}\sim \omega_0/\ln(1/h)$,
then Eq. (10) reduces to the following rough asymptotic equation for
$\Delta E_{NR}$:

\begin{eqnarray}
&&\Delta E_{NR}\sim V_0(E=E_s\pm\Delta E_{NR})h\ln(1/h), \\
%\quad\quad
&&h\rightarrow 0 .\nonumber
\end{eqnarray}

The asymptotic solution of Eq. (11) essentially depends on
$V_0(E_s\pm\Delta E_{NR})$ as a function of $\Delta E_{NR}$. In this
context, all systems can be divided in the following two types.

{\bf Type I}. The separatrix of the unperturbed system has {\it two
or more} saddles while the relevant Fourier coefficient
$V^{(L)}\equiv V^{(L)}(E,\psi)$ possesses {\it different} values on
adjacent saddles. Given that, for $E\rightarrow E_s$, the system
stays most of time near one of the saddles, the coefficient
$V^{(L)}(E\rightarrow E_s,\psi)$ as a function of $\psi$ is nearly a
\lq\lq square wave\rq\rq: it oscillates between the values at the
different saddles. The relevant $K$ is typically odd and, therefore,
$V_0(E\rightarrow E_s)$ approaches a well defined non-zero value.
Substituting it in Eq. (11), we conclude that

\begin{equation}
\Delta E_{NR}\propto h\ln(1/h), \quad\quad h\rightarrow 0 .
\end{equation}

{\bf Type II}. Either (i) the separatrix of the unperturbed system
has a {\it single saddle}, or (ii) it has more than one saddle but
the perturbation coefficient $V^{(L)}$ is {\it identical} for all
saddles. Then $V^{(L)}(E\rightarrow E_s,\psi)$, as a periodic
function of $\psi$, significantly differs from its value at the
saddle(s) only during a small part of the period in $\psi$: this
part is $\sim \omega(E)/\omega_0\sim 1/\ln(1/|E_s-E|)$. Hence,
$V_0(E_s\pm\Delta E_{NR})\propto 1/\ln(1/\Delta E_{NR})$.
Substituting this value in Eq. (11), we conclude that

\begin{equation}
\Delta E_{NR}\propto h, \quad\quad h\rightarrow 0 .
\end{equation}

Thus, for systems of type I, the maximal width of the SM chaotic layer is proportional to $h$ times a logarithmically
large factor $\propto\ln(1/h)$ while, for systems of type II,
%the maximal width
it
is proportional to $h$ times a numerical factor.

As shown below, the variation of energy in between the instants
relevant to the SM is $\sim h$, which thus it is much less than $\Delta E_{NR}$
(12) for the systems of type I and it is of the same order of $\Delta
E_{NR}$
(13) for the systems of type II. Therefore,
one may expect that the maximal width of the layer for the original Hamiltonian
system (1), $\Delta E^{(\pm)}$, is at least roughly approximated by that for
the SM, $\Delta E_{sm}^{(\pm)}$, so that the above
classification of systems is relevant to $\Delta E^{(\pm)}$ too. This is confirmed
both by the numerical integration of equations of motion of the original
Hamiltonian system and by the more accurate theory presented in the
next two sections.

\section{Asymptotic theory for systems of type I.}

For the sake of clarity, we
%shall
consider a concrete example of
type I, while the generalization is
straightforward.

Let us consider an archetypal example: the ac-driven pendulum
(sometimes called as a pendulum subject to a dipole time-periodic
perturbation) \cite{Zaslavsky:1991,prl2005,13_prime}:

\begin{eqnarray}
&& H=H_0+hV,
\\
&&  H_0=\frac{p^2}{2}-\cos(q), \quad\quad V=-q\cos(\omega_ft),
\quad\quad h\ll 1. \nonumber
\end{eqnarray}

Fig. 2 presents the results of computer simulations (i.e. of a
%direct
numerical integration of the equations of motion) for a few
values of $h$ and several values of $\omega_f$. It shows that:
1) the function $\Delta E^{(-)}(\omega_f)$ indeed possesses sharp peaks;
their height greatly exceed the estimates by the
heuristic \cite{Zaslavsky:1991}, adiabatic \cite{E&E:1991}
and mathematical moderate-frequency \cite{treschev}  theories (see the inset);
2) as
predicted by the rough estimates in Sec. III, the 1st peak of $\Delta
E^{(-)}(\omega_f)$ shifts to smaller values of $\omega_f$ while its
magnitude grows, as $h$ decreases. Below, we develop the
leading-order asymptotic theory and compare it with results of the
simulations.
%numerical integration of equations of motion.

\begin{figure}[tb]
\includegraphics*[width = 8.5 cm]{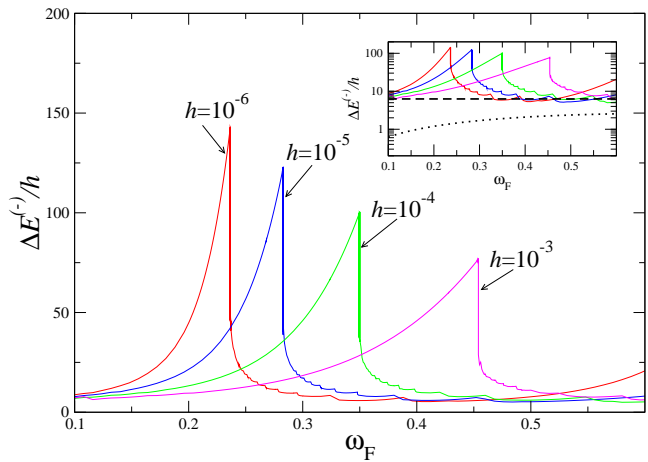}
\vskip 0.2 cm\caption {Computer simulations for the ac driven
pendulum (14) (an archetypal example of type I): the deviation
$\Delta E^{(-)}$ of the lower boundary of the chaotic layer from the
separatrix, normalized by the perturbation amplitude $h$, as a
function of the perturbation frequency $\omega_f$, for various $h$.
The inset presents the same data but in logarithmic vertical scale and with the estimates
by the heuristic \cite{Zaslavsky:1991}, adiabatic \cite{E&E:1991} and mathematical moderate-frequency \cite{treschev}
theories: the heuristic estimate is shown by the dotted line \cite{29_prime} while the adiabatic and moderate-frequency estimates are shown by the dashed line \cite{29_prime_prime}. The inset explicitly
shows that the simulation results exceed the estimates by the former theories by 1 or 2 orders of magnitude, for a wide range of frequencies.}
\end{figure}

\begin{figure}[htb]
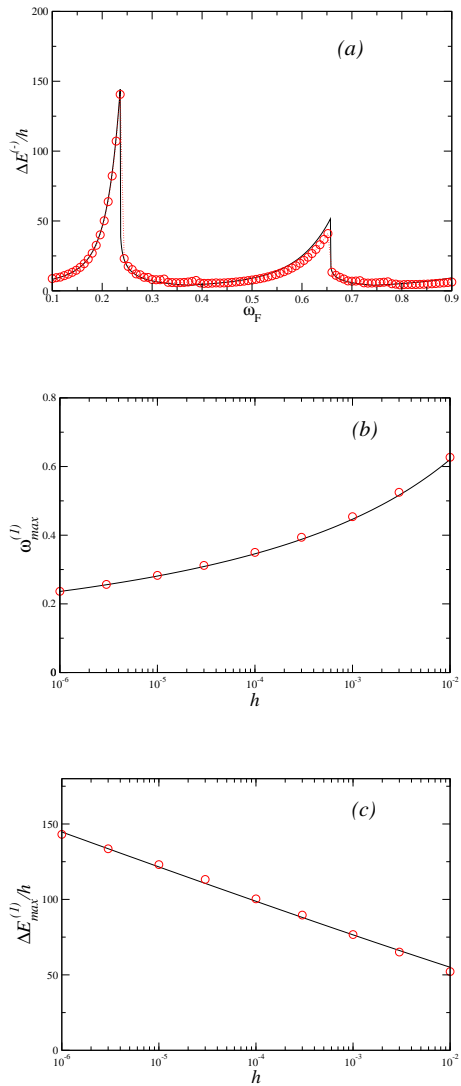

\includegraphics*[width = 6 cm]{Fig3a.eps}
\vskip 0.9 cm
\includegraphics*[width = 6 cm]{Fig3b.eps} \vskip 0.9
cm
\includegraphics*[width = 6 cm]{Fig3c.eps}
\vskip 0.2 cm \caption {An archetypal example of type I: ac-driven
pendulum (14). Comparison of theory (solid lines) and simulations
(circles):
 (a) the deviation $\Delta E^{(-)}(\omega_f)$ of the lower boundary of the chaotic
layer from the separatrix,
 normalized by the perturbation amplitude $h$, as a function of the perturbation frequency
 $\omega_f$, for $h=10^{-6}$; the theory is by Eqs. (26), (31), (32), (38), (39) and (41). (b)
 the frequency of the 1st maximum in $\Delta E^{(-)}(\omega_f)$
 as a function of $h$; the theory is by Eq. (26). (c) the 1st maximum in $\Delta E^{(-)}(\omega_f)/h$
 as a function of $h$; the theory is by Eqs. (34) and (26).} \label{abb:fig3}
\end{figure}

Before moving on, we note that the
SM
(approximated in
the relevant case by the nonlinear resonance dynamics) considers
states of the system only at discrete instants. Apart from the
variation of energy within the SM dynamics, the variation of energy
in the Hamiltonian system occurs also in between the instants
relevant to the SM. Given that $\omega_f\ll 1$, this latter
variation may be considered in adiabatic approximation and it is of
the order of $h$ \cite{E&E:1991,shevchenko}. As follows from the
rough estimates above and from the accurate consideration below, the
variation of energy within the
SM
dynamics for systems
of type I is logarithmically larger i.e. larger by the factor $\ln(1/h)$.
Therefore, the variation of
energy in between the instants relevant to the SM may be neglected
at the leading-order approximation for systems of type I (the parameter of
smallness of the asymptotic theory is $1/\ln(1/h)$): $\Delta
E^{(-)}\simeq \Delta E^{(-)}_{sm}$. For the sake of notational compactness,
we shall omit the subscript "$sm$" further in this
section.

For the system (14), the separatrix energy is equal to 1, while the
asymptotic (for $E\rightarrow E_s$) dependence $\omega(E)$ is
\cite{Zaslavsky:1991}:

\begin{eqnarray}
&& \omega(E)\simeq \frac{\pi}{\ln(32/|E_s-E|)},
\\
&&  E_s=1, \quad\quad |E_s-E|\ll 1. \nonumber
\end{eqnarray}

Let us consider the range of energies below $E_s$ (the range above
$E_s$ may be considered analogously) and assume that $\omega_f$ is
close to one of the odd multiples of $\omega_r^{(-)}$. The
%in accordance with the averaging method \cite{bogmit},
nonlinear resonance dynamics of the slow variables in the range of the
approximately resonant energies may be described as follows
\cite{pre2008,PR} (cf. also
\cite{Chirikov:79,lichtenberg_lieberman,Zaslavsky:1991,zaslavsky:1998,zaslavsky:2005,abdullaev}):

\begin{eqnarray}
&& \frac{{\rm d}I}{{\rm
d}t}=-\frac{\partial{\tilde{H}}(I,{\tilde{\psi}})}{\partial{\tilde{\psi}}},
\quad\quad \frac{{\rm d}{\tilde{\psi}}}{{\rm
d}t}=\frac{\partial{\tilde{H}}(I,{\tilde{\psi}})}{\partial I},
\\
&& \tilde{H}(I,\tilde{\psi})=\int_{I(E_s)}^{I}{\rm d}\tilde{I}\;
(n\omega-\omega_f)\;-\; nhq_n\cos(\tilde{\psi}) \nonumber
\\
&& \quad\; \equiv\; n(E-E_s)-\omega_f(I-I(E_s))\;-\;
nhq_n\cos(\tilde{\psi})\;, \nonumber
\\
&& I \equiv I(E) = \int_{E_{\rm min}}^E \frac{{\rm
d}\tilde{E}}{\omega(\tilde{E})}, \quad\quad  E \equiv H_0(p,q),
\nonumber
\\
&& \tilde{\psi}=n\psi-\omega_ft, \quad\quad \nonumber
\\
&& \psi= \pi+{\rm sign}(p)\omega(E)\int^q_{q_{\rm min}(E)}\frac{{\rm
d}\tilde{q}}{\sqrt{2(E-U(\tilde{q}))}}+2\pi l, \nonumber
\\
&& q_n\equiv q_n(E)= \frac{1}{2\pi}\int_0^{2\pi} \!\!\!\!\! {\rm
d}\psi \; q(E,\psi)\cos(n\psi) ,
%\quad i\equiv \sqrt{-1},
\nonumber
\\
&& |n\omega-\omega_f|\ll\omega,\quad\quad n\equiv 2j-1, \quad\quad
j=1,2,3,\ldots, \nonumber
\end{eqnarray}

\noindent where $I$ and $\psi$ are the canonical variables
action and angle respectively
\cite{Chirikov:79,lichtenberg_lieberman,Zaslavsky:1991,zaslavsky:1998,zaslavsky:2005,abdullaev};
$E_{\rm min}$ is the minimal energy over all $q,p$,
$ E \equiv H_0(p,q)$;
%$\omega\equiv \omega(E) =d H_0/d I$ and
$q_{\rm min}(E)$ is the minimal coordinate of the conservative
motion with a given value of energy $E$; $l$ is the number of right
turning points in the trajectory $[q(\tau)]$ of the conservative
motion with energy $E$ and given initial state $(q_0,p_0)$.

The resonance Hamiltonian $\tilde{H}(I,\tilde{\psi})$
is obtained from the original Hamiltonian $H$
transforming to action-angle variables $I-\psi$, with a further
multiplication by $n$; extracting the term $\omega_fI$ (that
corresponds to the transformation $\psi\rightarrow\tilde{\psi}\equiv
n\psi-\omega_ft$); and neglecting all the fast-oscillating terms
(their effect on the dynamics of slow variables is small:
%\cite{bogmit}
see the estimate of the corrections in Sec. VI below) i.e.
keeping only the
resonance term in the double Fourier expansion of the perturbation.

Let us derive the asymptotic expression for $I(E)$, substituting the
asymptotic expression (15) for $\omega(E)$ into the definition of
$I(E)$ (16) and carrying out the integration:

\begin{equation}
I(E)\simeq I(E_s)-\frac{E_s-E}{\pi}\left ( \ln \left
(\frac{32}{E_s-E} \right )+1 \right ).
\end{equation}

As for the asymptotic value $q_n(E\rightarrow E_s)$, it is easy to
see that $q(E\rightarrow E_s,\psi)$, as a function of $\psi$,
asymptotically approaches a \lq\lq square wave\rq\rq, oscillating
between $-\pi$ and $\pi$, so that, for sufficiently small $j$,

\begin{eqnarray}
&& q_{2j-1}(E\rightarrow E_s)\simeq (-1)^{j+1}\frac{2}{2j-1},\\
%\quad\quad
&& q_{2j}=0,\nonumber\\
&& j=1,2,...\ll \frac{\pi}{2\omega(E)}.\nonumber
\end{eqnarray}

The next issue is the analysis of the phase space of the resonant
Hamiltonian (16). Substituting $\tilde{H}$ (16) into the equations
of motion (16), it is easy to see that their stationary points have
the following values of the slow angle

\begin{equation}
\tilde{\psi}_+=\pi, \quad\quad \tilde{\psi}_-=0,
\end{equation}

\noindent while the corresponding action is determined by the
equation

\begin{equation}
n\omega-\omega_f\mp nh\frac{{\rm d}q_n}{{\rm d}I}=0, \quad\quad
n\equiv 2j-1,
\end{equation}

\noindent where the sign \lq\lq $\mp$\rq\rq corresponds to
$\tilde{\psi}_{\mp}$ (19).

As usual (cf.
\cite{Chirikov:79,lichtenberg_lieberman,Zaslavsky:1991,zaslavsky:1998,zaslavsky:2005,abdullaev,pre2008,PR}),
the term $\propto h$ in (20) may be neglected in the leading-order
approximation, and Eq. (20) reduces to the resonance condition

\begin{equation}
(2j-1)\omega(E_r^{(j)})=\omega_f,
\end{equation}

\noindent the lowest-order solution of which is

\begin{equation}
E_s-E_r^{(j)}\simeq 32\exp\left(-\frac{(2j-1)\pi}{\omega_f}\right).
\end{equation}

Eqs. (19) and (22) together with (17) explicitly determine the
elliptic and hyperbolic points of the Hamiltonian (16). The
hyperbolic point is often called \lq\lq saddle'' and corresponds to
$\tilde{\psi}_+$ or $\tilde{\psi}_-$ in (19) for even or odd $j$ respectively. The saddle
point generates the resonance separatrix. Using the
asymptotic relations (17) and (18), we obtain that the resonance Hamiltonian (16)
takes the following asymptotic value in the saddle:

\begin{eqnarray}
&&\tilde{H}_{saddle}\simeq \frac{E_s-E_r^{(j)}}{\pi}\omega_f-2h\nonumber\\
&&\quad\quad\quad \simeq  \frac{\omega_f}{\pi}32\exp\left(-\frac{\pi
(2j-1)}{\omega_f}\right)-2h.
\end{eqnarray}

\noindent The second asymptotic equality in (23) takes into account
the relation (22).

As explained in Sec. II above, $\Delta E^{(-)}(\omega_f)$
possesses a local maximum at $\omega_f$ for which the resonance
separatrix is tangent to the lower GSS curve (Fig. 1(a)). For the
relevant frequency range $\omega_f\rightarrow 0$, the separatrix
split (which represents the maximum deviation of the energy along the GSS curve
from $E_s$)
approaches the following
value \cite{Zaslavsky:1991}, in the asymptotic limit $h\rightarrow 0$

\begin{equation}
\delta\simeq  2\pi h, \quad\quad \omega_f\ll 1.
\end{equation}

\noindent As it is shown further down, the variation of energy along the
relevant resonance trajectories is much larger. Therefore, in the
leading-order approximation, the GSS curve may be simply replaced by
the separatrix of the unperturbed system i.e. by the horizontal line
$E=E_s$ or, equivalently, $I=I(E_s)$. Then the tangency occurs at $\tilde{\psi}$ shifted from the
saddle by $\pi$, so that the condition of tangency is written as

\begin{equation}
\tilde{H}_{saddle}=\tilde{H}(I=I(E_s),\tilde{\psi}=\tilde{\psi}_{saddle}+\pi)\equiv
2h.
\end{equation}

Substituting here $\tilde{H}_{saddle}$ (23), we finally obtain the
following transcendental equation for $\omega_{\max}^{(j)}$:

\begin{equation}
x\exp(x)=\frac{8(2j-1)}{h},\quad\quad x\equiv
\frac{(2j-1)\pi}{\omega_{\max}^{(j)}}.
\end{equation}

\noindent Fig. 3(b) demonstrates the excellent agreement between Eq.
(26) and the results of simulations for the Hamiltonian system in a
wide range of $h$.

In the asymptotic limit $h\rightarrow 0$, the lowest-order explicit
solution of Eq. (26) is

\begin{equation}
\omega_{\max}^{(j)}\simeq
\frac{(2j-1)\pi}{\ln\left(\frac{8(2j-1)}{h}\right)}, \quad\quad
j=1,2,...\ll \ln\left(\frac{1}{h}\right).
\end{equation}

As follows from Eq. (26), the value of $E_s-E_r^{(j)}$ (22) for
$\omega_f=\omega_{\max}^{(j)}$ is

\begin{equation}
E_s-E_r^{(j)}(\omega_f=\omega_{\max}^{(j)})=\frac{4\pi
h}{\omega_{\max}^{(j)}}.
\end{equation}

Its leading-order expression is:

\begin{eqnarray}
&& E_s-E_r^{(j)}(\omega_f=\omega_{\max}^{(j)})\simeq \frac{4
h}{2j-1}\ln\left(\frac{8(2j-1)}{h}\right),\nonumber\\
%\quad\quad
&& h\rightarrow 0.
\end{eqnarray}

If $\omega_f\leq\omega_{\max}^{(j)}$, then, in the chaotic layer, the
largest
%maximum
deviation of
energy from the separatrix value corresponds to the
minimum energy $E_{\min}^{(j)}$ on the nonlinear resonance
separatrix (Fig. 1(a,b)), which occurs at $\tilde{\psi}$ shifted by
$\pi$ from the saddle. The condition of equality of $\tilde{H}$
at the saddle and at the minimum of the resonance separatrix is
written as

\begin{equation}
\tilde{H}_{saddle}=\tilde{H}(I(E_{\min}^{(j)}),\tilde{\psi}_{saddle}+\pi).
\end{equation}

Let us seek its asymptotic solution in the form

\begin{eqnarray}
&& E_s-E_{\min}^{(j)}\equiv\Delta E_l^{(j)}=(1+y)
(E_s-E_{r}^{(j)})\nonumber\\
&&\quad\quad\quad\quad\quad\quad\quad\quad\quad\simeq
(1+y)32\exp\left(-\frac{\pi (2j-1)}{\omega_f}\right),
\nonumber\\
&& y\stackrel{>}{\sim}1.
\end{eqnarray}

Substituting (31) and (23) into Eq. (30), we obtain for $y$ the
following transcendental equation:

\begin{eqnarray}
&& (1+y)\ln(1+y)-y=\frac{h}{8(2j-1)}x_f\exp(x_f),
\\
&&  x_f\equiv \frac{\pi (2j-1)}{\omega_f}, \quad\quad
\omega_f\leq\omega_{\max}^{(j)},\quad\quad y>0, \nonumber
\end{eqnarray}

\noindent where $\omega_{\max}^{(j)}$ is given by Eq. (26).

Eqs. (31) and (32) describe the left wing of the $j$-th peak of
$\Delta E^{(-)}(\omega_f)$. Fig. 3(a) demonstrates the good
agreement between our analytic theory and simulations for the
Hamiltonian system.

As follows from Eq. (26), Eq. (32) for
$\omega_f=\omega_{\max}^{(j)}$ reduces to the relation $\ln(1+y)=1$,
i.e.

\begin{equation}
1+y(\omega_{\max}^{(j)}) = {\rm e}.
\end{equation}

As follows from Eqs. (33), (31) and (28), the maximum for a given
peak is:

\begin{equation}
\Delta E^{(j)}_{\max}\equiv
E_s-E^{(j)}_{\min}(\omega_{\max}^{(j)})=\frac{4\pi{\rm
e}h}{\omega_{\max}^{(j)}}.
\end{equation}

Fig. 3(c) shows the excellent agreement of this expression with the
results of simulations for the Hamiltonian system in a wide range of
$h$.

The leading-order expression for $\Delta E^{(j)}_{\max}$ is:

\begin{equation}
\Delta E^{(j)}_{\max}\simeq\frac{4{\rm
e}h}{2j-1}\ln(8(2j-1)/h),\quad\quad h\rightarrow 0,
\end{equation}

\noindent that confirms the rough estimate (12).

As $\omega_f$ decreases, $y$ increases exponentially sharply, as
follows from Eq. (32). In order to understand how $\Delta E_l^{(j)}$
decreases upon decreasing $\omega_f$, it is convenient to
rewrite Eq. (31) expressing the exponent by means of Eq. (32):

\begin{equation}
\Delta E_l^{(j)}(\omega_f)=\frac{4\pi
h}{\omega_f(\ln(1+y)-y/(1+y))}.
\end{equation}

\noindent It follows from Eqs. (32) and (36) that $\Delta E_l^{(j)}$
decreases power-like rather than exponentially when $\omega_f$
is decreased. In particular,
$\Delta E_l^{(j)}\propto 1/(\omega_{\max}^{(j)}-\omega_f)$
for the far part of the wing.

As for the right wing of the peak, i.e. for
$\omega_f>\omega_{\max}^{(j)}$, over the chaotic layer the
largest
%maximum
deviation of
energy from the separatrix value corresponds to the minimum of
the resonance trajectory tangent to the GSS curve (Fig. 1(c)). The
value of $\tilde{\psi}$ in the minimum coincides with
$\tilde{\psi}_{saddle}$. In the leading-order approximation, the GSS
curve may be replaced by the horizontal line $I=I(E_s)$, so that the
tangency occurs at $\tilde{\psi}=\tilde{\psi}_{saddle}+\pi$. Then
the energy at the minimum $E_{\min}^{(j)}$ can be found from the
equation

\begin{equation}
\tilde{H}(I(E_s),\tilde{\psi}_{saddle}+\pi)=\tilde{H}(I(E_{\min}^{(j)}),\tilde{\psi}_{saddle})
\end{equation}

Let us seek its asymptotic solution in the form

\begin{eqnarray}
&& E_s-E_{\min}^{(j)}\equiv\Delta E_r^{(j)}= z
(E_s-E_{r}^{(j)})\nonumber\\
&&\quad\quad\quad\quad\quad\quad\simeq z32\exp\left(
-\frac{\pi(2j-1)}{\omega_f}\right)
\nonumber\\
&& 0<z< 1, \quad\quad z\sim 1.
\end{eqnarray}

Substituting (38) into Eq. (37), we obtain for $z$ the following
transcendental equation:

\begin{eqnarray}
&& z(1+\ln(1/z))=\frac{h}{8(2j-1)}x_f\exp(x_f)\\
&& x_f\equiv\frac{\pi(2j-1)}{\omega_f}, \quad\quad \omega_f>
\omega_{\max}^{(j)},\quad\quad 0<z<1, \nonumber
\end{eqnarray}

\noindent where $\omega_{\max}^{(j)}$ is given by Eq. (26).

Eqs. (38) and (39) describe the right wing of the $j$-th peak of
$\Delta E^{(-)}(\omega_f)$. Fig. 3(a) shows the good agreement
between our analytic theory and simulations.

As follows from Eq. (26), the solution of Eq. (39) for
$\omega_f\rightarrow\omega_{\max}^{(j)}$ is $z\rightarrow 1$, so
the right wing starts from the value given by Eq. (28) (or,
approximately, by Eq. (29)).

Expressing the exponent in (38) from (39), we obtain the following
equation

\begin{equation}
\Delta E_r^{(j)}(\omega_f)=\frac{4\pi h}{\omega_f(1+\ln(1/z))}.
\end{equation}

\noindent It follows from Eqs. (39) and (40) that $\Delta
E_r^{(j)}$ decreases power-like rather than exponentially for
increasing $\omega_f$.
In particular,
$\Delta E_r^{(j)}\propto 1/(\omega_f-\omega_{\max}^{(j)})$
in the far part of the wing.

The further analysis of the asymptotic shape of the peak is done in
Sec. VII below.

Beyond the peaks, the function $\Delta E^{(-)}(\omega_f)$ is logarithmically small in comparison with the maxima of the peaks. The functions $\Delta E^{(j)}_l(\omega_f)$ and $\Delta E^{(j)}_r(\omega_f)$ in the ranges beyond the peaks are also logarithmically small. Hence, nearly any combination of the functions $\Delta E^{(j)}_r(\omega_f)$ and $\Delta E^{(j+1)}_l(\omega_f)$ which is close to $\Delta E^{(j)}_r(\omega_f)$ in the vicinity of $\omega_{\max}^{(j)}$ and to
$\Delta E^{(j+1)}_l(\omega_f)$ in the vicinity of $\omega_{\max}^{(j+1)}$ may be considered as an approximation of the function $\Delta E^{(-)}(\omega_f)$ with a logarithmic accuracy with respect to the maxima of the peaks, $\Delta E_{\max}^{(j)}$ and $\Delta E_{\max}^{(j+1)}$, in the whole range  $[\omega_{\max}^{(j)},\omega_{\max}^{(j+1)}]$. One of the easiest combinations is the following:

\begin{eqnarray}
&& \Delta E^{(-)}(\omega_f)=\Delta E^{(1)}_l(\omega_f) \quad\quad {\rm for}\quad\omega_f<
\omega_{\max}^{(1)},
\nonumber\\
&&\Delta E^{(-)}(\omega_f)=\max\{\Delta E^{(j)}_r(\omega_f),\Delta
E^{(j+1)}_l(\omega_f) \}\nonumber\\&&\quad\quad\quad\quad\quad\quad
{\rm for}\quad\omega_{\max}^{(j)}<\omega_f<
\omega_{\max}^{(j+1)}, \nonumber\\
&&j=1,2,...\ll \frac{\pi}{2\omega_{\max}^{(1)}}.
\end{eqnarray}

\noindent We used this function in Fig. 3(a), and the analogous
combination will be also used in the other cases.

In fact, the theory may be generalized in such a way that Eq. (41)
would well approximate $\Delta E^{(-)}(\omega_f)$ in the ranges
far beyond the peaks with a logarithmic accuracy even with respect
to $\Delta E^{(-)}(\omega_f)$ itself rather than to $\Delta E_{\max}^{(j)}$
only  (cf. the next section). However, we do not do this in the present case,
being interested primarily in the leading-order description of the peaks.

Finally, we demonstrate in Fig. 4 that the lowest-order theory
describes quite well the layer boundaries even in the Poincar\'{e}
section rather than only in energy/action.

\begin{figure}[htb]
\includegraphics*[width = 6.3 cm]{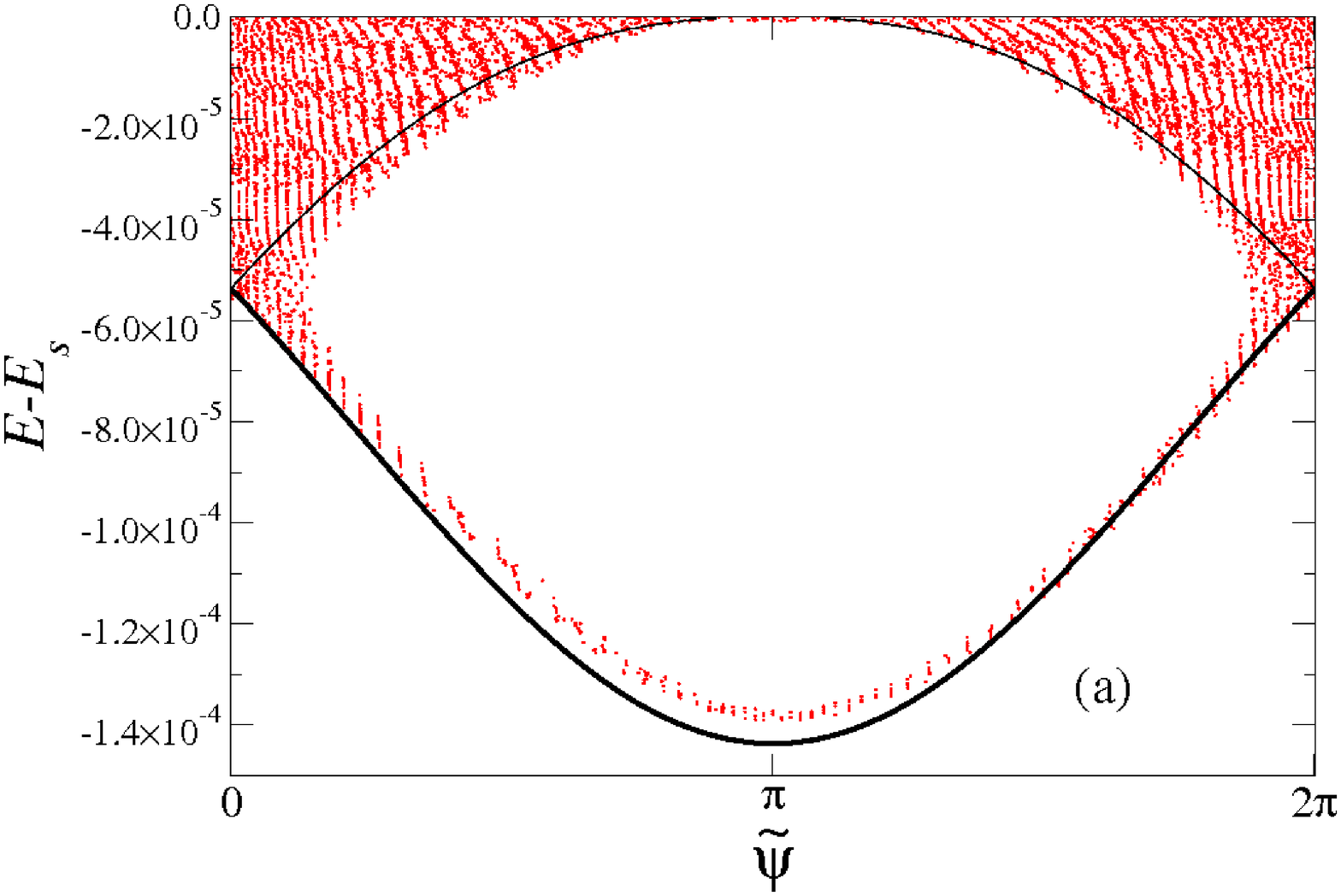}
\vskip 0.2 cm
\includegraphics*[width = 6.3 cm]{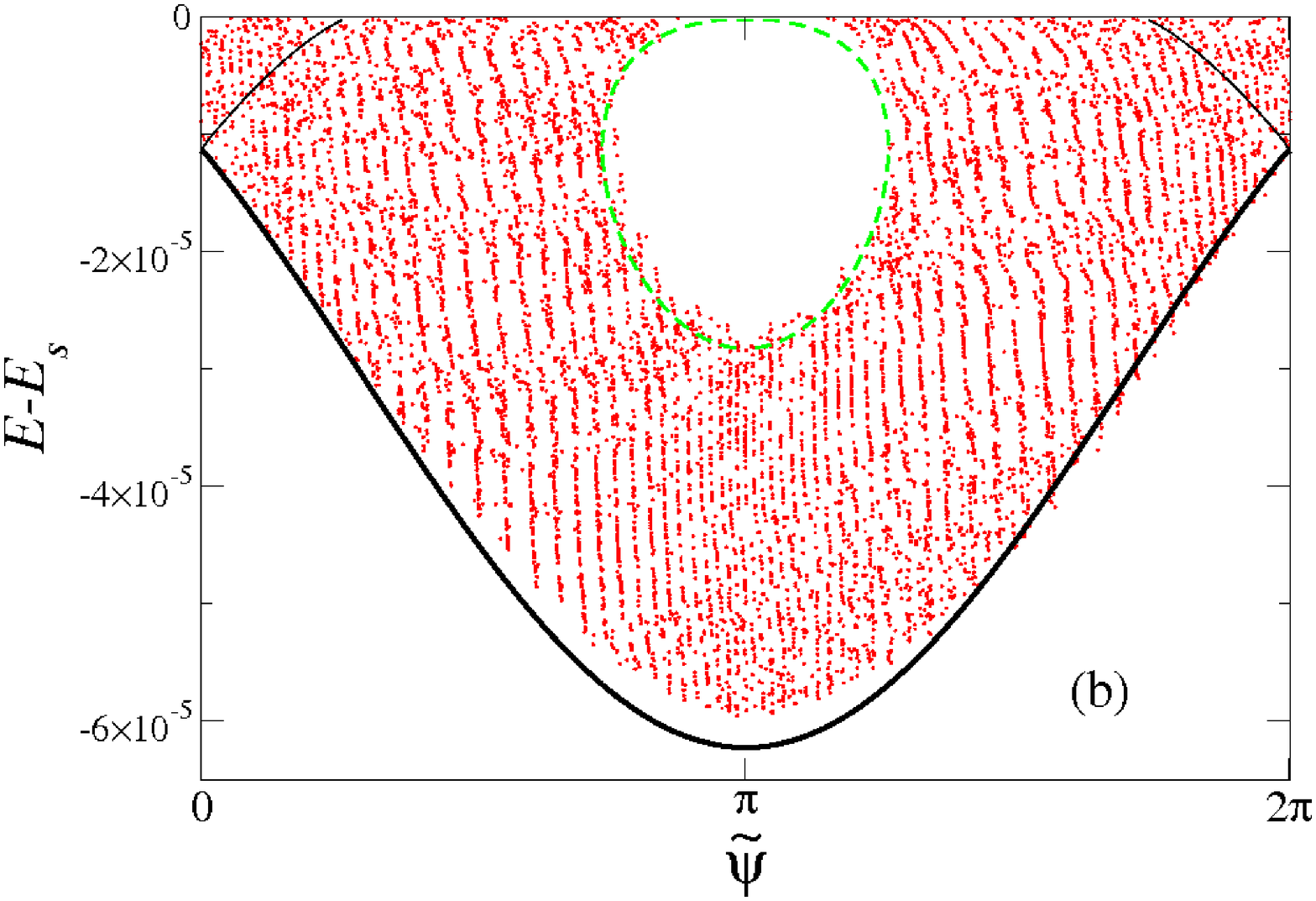}
\vskip 0.2 cm
\includegraphics*[width = 6.3 cm]{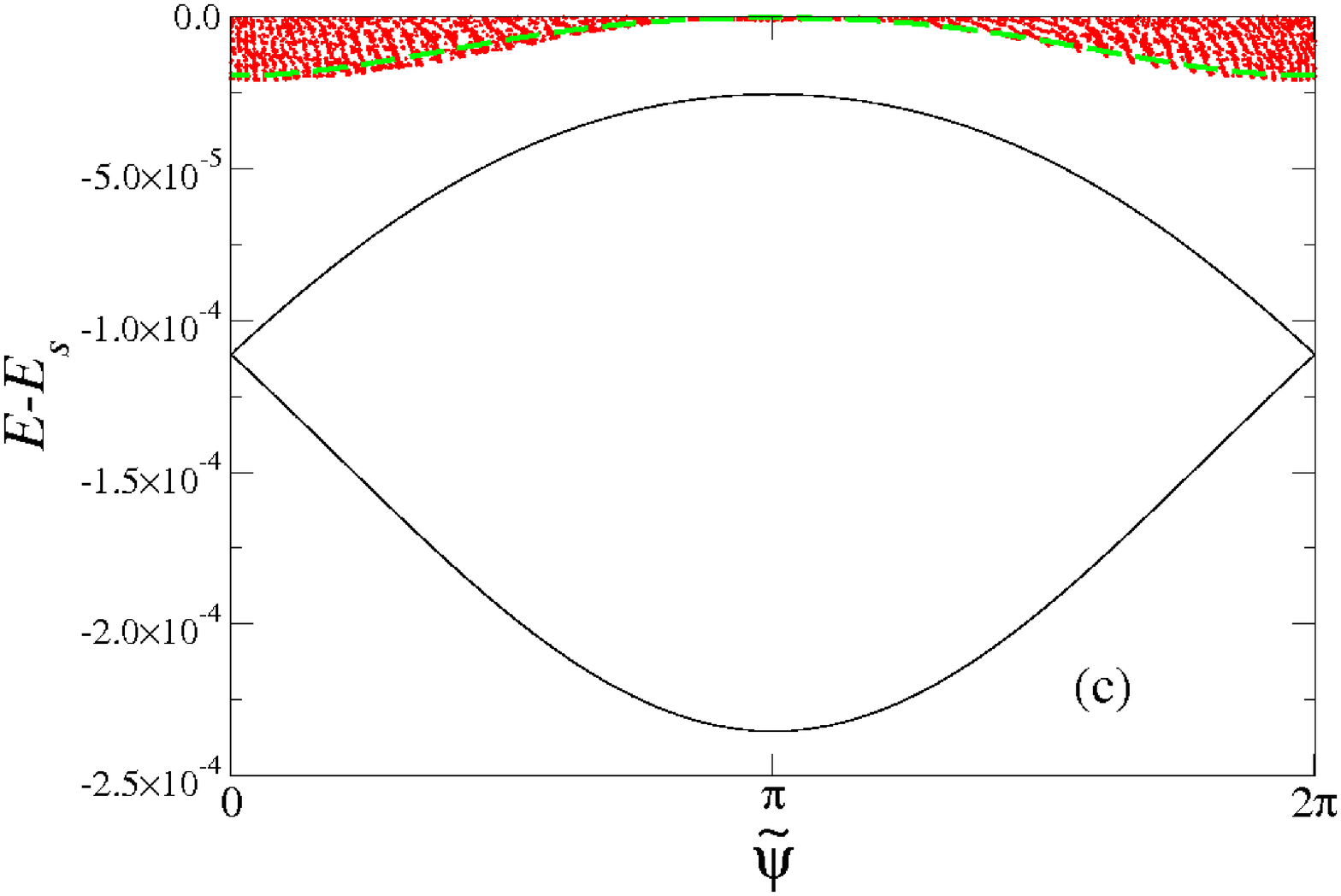}
\vskip 0.2 cm \caption {A few characteristic Poincar\'{e} sections
in the $2\pi$-interval of the energy-angle plane for the system (14)
with $h=10^{-6}$ and $\omega_f$ equal to: (a) 0.236 (maximum), (b)
0.21 (left wing), (c) 0.25 (right wing). Results of the numerical
integration of the equations of motion for the original Hamiltonian
(14) are shown by red dots. The NR separatrix calculated in the leading-order approximation
(i.e. by the integration of the resonant equations of motion (16) in which
$\omega(E)$, $I(E)$ and $q_1(E)$ are approximated by the explicit
formulas (15), (17) and (18) respectively)
is drawn by the black
solid line. The NR trajectory (calculated in the leading-order approximation) tangent to the line $E=E_s$ is
drawn by the blue dashed line. The outer boundary (marked by a
thicker line) is approximated by: the lower part of the NR
separatrix in the cases (a) and (b), and by the tangent NR
trajectory in the case (c) The boundary of the island of stability
in the cases (a) and (b) is approximated by the tangent NR
trajectory (which coincides in the case (a) with the NR
separatrix).} \label{abc:fig4}
\end{figure}

\section{Asymptotic theory for systems of type II.}

We shall consider two characteristic examples of type II,
corresponding to the classification given in Sec. III. As an
example of the system where the separatrix of the unperturbed
system possesses a single saddle, we shall consider the ac-driven
Duffing oscillator \cite{abdullaev,gelfreich,treschev,soskin2000}. As an
example of the system where the separatrix possesses more than one
saddle while the perturbation takes equal values at the saddles,
we shall consider the pendulum with an oscillating suspension
point \cite{abdullaev,gelfreich,treschev,shevchenko:1998,shevchenko}. The
treatment of these cases is similar in many respects to the one
presented in Sec. IV above. So, we present it in less details,
emphasizing the differences.

\subsection{AC-driven Duffing oscillator.}

Consider the following archetypal Hamiltonian
\cite{abdullaev,gelfreich,treschev,soskin2000}:

\begin{eqnarray}
&& H=H_0+hV,
\\
&&  H_0=\frac{p^2}{2}-\frac{q^2}{2}+\frac{q^4}{4}, \quad\quad
V=-q\cos(\omega_ft), \quad\quad h\ll 1. \nonumber
\end{eqnarray}

The asymptotic dependence of $\omega(E)$ on $E$ for $E$ below the
separatrix energy $E_s=0$ is the following
\cite{abdullaev,physica1985}

\begin{eqnarray}
&& \omega(E)\simeq \frac{2\pi}{\ln(16/(E_s-E))},
\\
&&  E_s=0, \quad\quad 0<E_s-E\ll 1. \nonumber
\end{eqnarray}

Correspondingly, the resonance values of energies (determined by the
condition analogous to (21)) are

\begin{equation}
E_s-E_r^{(j)}=16\exp\left(-\frac{2\pi j}{\omega_f}\right),
\quad\quad j=1,2,3,...
\end{equation}

The asymptotic dependence of $I(E)$ is

\begin{equation}
I(E)\simeq I(E_s)-\frac{E_s-E}{2\pi}\left ( \ln \left
(\frac{16}{E_s-E} \right )+1 \right ).
\end{equation}

The nonlinear resonance dynamics is described by the resonance
Hamiltonian $\tilde{H}$ which is identical to Eq. (16) in form.
Obviously, the actual dependencies $\omega(E)$ and $I(E)$ are
given by Eq. (43) and (45) respectively. The most important
difference is in $q_j(E)$: instead of a non-zero value (see (18)),
it approaches 0 as $E\rightarrow E_s$. Namely, it is $\propto
\omega(E)$ \cite{abdullaev,physica1985}:

\begin{equation}
q_j(E)\simeq \frac{1}{\sqrt{2}}\omega(E),\quad\quad j=1,2,...\ll
\frac{\pi}{\omega(E)},
\end{equation}

\noindent i.e. $q_j$ is much smaller than in systems of type I
(cf. (18)). Due to this, the resonance is \lq\lq weaker\rq\rq. At
the same time, the separatrix split $\delta$ is also smaller, namely $\sim
h\omega_f$ (cf. \cite{pre2008}) rather than $\sim
h$ as for the systems of type I. That is why the separatrix chaotic layer is still dominated by the resonance dynamics while
the matching of the
separatrix map and nonlinear resonance dynamics is still valid in
the asymptotic limit $h\rightarrow 0$ \cite{pre2008}.

Similarly to the previous section, we find the value of $\tilde{H}$ in
the saddle in the leading-order approximation \cite{27_prime}:

\begin{equation}
\tilde{H}_{saddle}\simeq \omega_f\left(
\frac{E_s-E_r^{(j)}}{2\pi}-\frac{h}{\sqrt{2}} \right),
\end{equation}

\noindent where $E_s-E_r^{(j)}$ is given in (44).

As before, the maximum width of the layer corresponds to $\omega_f$, for
which the resonance separatrix is tangent to the GSS curve (Fig.
1(a)). It can be shown \cite{pre2008} that the angle of tangency
asymptotically approaches $\tilde{\psi}_{saddle}+\pi= \pi$ while the
energy still lies in the resonance range, where $\omega(E)\approx
\omega_{r}^{(-)}\approx \omega_f/j$. Using the expressions for $\tilde{H}(E,\tilde{\psi})$ (cf. (16)), $I(E)$ (45), $q_j(E)$ (46), and taking into account that in the tangency $E <\delta\sim h\omega_f\ll h$, the value of
$\tilde{H}$ at the
tangency reads, in the leading-order approximation,
\begin{equation}
\tilde{H}_{tangency}\simeq \omega_f \frac{h}{\sqrt{2}}.
\end{equation}

Allowing for Eqs. (47) and (48), the condition for the maximum,
$\tilde{H}_{saddle}=\tilde{H}_{tangency}$, reduces to

\begin{equation}
E_s-E_r^{(j)}(\omega_{\max}^{(j)})\simeq 2\pi \sqrt{2}h.
\end{equation}

Thus, these values $E_s-E_r^{(j)}$ are logarithmically smaller than
the corresponding values (28) for systems of type I.

The values of $\omega_f$ corresponding to the maxima of the peaks in
$\Delta E^{(-)}(\omega_f)$ are readily obtained from (49) and
(44):

\begin{equation}
\omega_{\max}^{(j)}\simeq \frac{2\pi j}{\ln(4\sqrt{2}/(\pi
h))},\quad\quad j=1,2,...\ll\ln(1/h).
\end{equation}

The derivation of the shape of the peaks for the chaotic layer of
the separatrix map in the leading order, i.e. within the nonlinear
resonance (NR) approximation, is similar to that for type I. So,
we present only the results, marking them with the subscript
\lq\lq $NR$\rq\rq.

The left wing of the $j$th peak of $\Delta E^{(-)}_{NR}(\omega_f)$
is described by the function

\begin{eqnarray}
&&\Delta E^{(j)}_{l,NR}(\omega_f)=16(1+y)\exp\left(-\frac{2\pi
j}{\omega_f}\right)
\\&&\quad\quad\quad\quad\quad\quad\equiv\frac{2\pi\sqrt{2}
h}{\ln(1+y)-y/(1+y)},\quad\quad
\omega_f\leq\omega_{\max}^{(j)},\nonumber
\end{eqnarray}

\noindent where $y$ is the positive solution of the transcendental
equation

\begin{equation}
(1+y)\ln(1+y)-y=\frac{\pi h}{4\sqrt{2}}\exp\left(\frac{2\pi
j}{\omega_f}\right),\quad\quad y>0.
\end{equation}

Similarly to the type I case, $1+y(\omega_{\max}^{(j)})={\rm e}$, so
that

\begin{equation}
\Delta E^{(j)}_{\max,NR}={\rm e}
(E_s-E_r^{(j)}(\omega_{\max}^{(j)}))\simeq 2\pi{\rm e}\sqrt{2} h.
\end{equation}

Eq. (53) confirms the rough estimate (13).

The right wing of the peak is described by the function

\begin{eqnarray}
&&\Delta E^{(j)}_{r,NR}(\omega_f)=16z\exp\left(-\frac{2\pi
j}{\omega_f}\right)\nonumber\\ &&\quad\quad
\equiv\frac{2\pi\sqrt{2} h}{1+\ln(1/z)}, \quad\quad
\omega_f>\omega_{\max}^{(j)},
\end{eqnarray}

\noindent where $z<1$ is the solution of the transcendental equation

\begin{equation}
z(1+\ln(1/z))=\frac{\pi h}{4\sqrt{2}}\exp\left(\frac{2\pi
j}{\omega_f}\right),\quad\quad 0<z<1.
\end{equation}

\noindent Similarly to the type I case,
$z(\omega_f\rightarrow\omega_{\max}^{(j)})\rightarrow 1$.

As follows from Eqs. (49) and (53), the typical variation of
energy within the nonlinear resonance dynamics (that approximates
the separatrix map dynamics) is $\propto h$. For the Hamiltonian
system, the variation of energy in between the discrete instants
corresponding to the separatrix map
\cite{Zaslavsky:1991,zaslavsky:1998,zaslavsky:2005,abdullaev,pre2008,vered} is also $\propto h$.
Therefore, unlike the case of type I, one needs to take it into
account even at the leading-order approximation. Let us consider
the right well of the Duffing potential (the results for the left
well are identical), and denote by $t_k$ the instant at which the
energy $E$ at a given $k$-th step of the separatrix map is taken:
it corresponds to the beginning of the $k$-th pulse of velocity
\cite{Zaslavsky:1991,pre2008} i.e. the corresponding $q$ is close
to a left turning point $q_{ltp}$ in the trajectory $[q(\tau)]$.
Let us also take into account that the relevant frequencies are
small so that the adiabatic approximation may be used. Thus, the
change of energy from $t_k$ up to a given instant $t$ during the
following pulse of velocity ($t-t_k\sim 1$) may be calculated as

\begin{eqnarray}
&&\Delta E =\int_{t_k}^{t}{\rm
d}\tau\dot{q}h\cos(\omega_f\tau)\simeq
h\cos(\omega_ft_k)\int_{t_k}^{t}{\rm
d}\tau\dot{q}\nonumber\\
&&\quad\quad=h\cos(\omega_ft_k)(q(t)-q_{ltp})
\end{eqnarray}

For the motion near the separatrix, the velocity pulse corresponds
approximately to $\psi=0$ (see the definition of $\psi$ (16)). Thus,
the corresponding slow angle is $\tilde{\psi}\equiv
j\psi-\omega_ft_k\simeq  -\omega_ft_k$.

For the left wing of the peak of $\Delta E^{(-)}(\omega_f)$
(including the maximum of the peak too), the boundary of the chaotic
layer of the separatrix map is formed by the lower part of the NR
separatrix. The minimum energy along this separatrix occurs at
$\tilde{\psi}=\pi$. Taking this into account, and also that
$\tilde{\psi}\simeq  -\omega_ft_k$, we conclude that
$\cos(\omega_ft_k)\simeq  -1$. So, $\Delta E\leq 0$, i.e. it does
lower the minimum energy of the layer of the Hamiltonian system. The
maximum lowering occurs at the right turning point $q_{rtp}$:

\begin{equation}
\max(|\Delta E|)\simeq  h(q_{rtp}-q_{ltp})=\sqrt{2}h.
\end{equation}

We conclude that the left wing of the $j$-th peak is described by
the following formula:

\begin{equation}
\Delta E_l^{(j)}(\omega_f)\simeq  \Delta
E_{l,NR}^{(j)}(\omega_f)+\sqrt{2}h,\quad\quad\omega_f\leq\omega_{\max}^{(j)},
\end{equation}

\noindent where $\Delta E_{l,NR}^{(j)}(\omega_f)$ is given by Eqs.
(51)-(52). In particular, the maximum of the peak is:

\begin{equation}
\Delta E^{(j)}_{\max}\simeq  (2\pi {\rm e}+1)\sqrt{2}h\approx 25.6h.
\end{equation}

For the right wing of the peak, the minimum energy of the layer of
the separatrix map occurs at $\tilde{\psi}$ coinciding with
$\tilde{\psi}_{saddle}$ (Fig. 1(c)) i.e. equal to 0. As a result,
$\cos(\omega_ft_k)\simeq 1$ and, hence, $\Delta E\geq 0$. So, this
variation cannot lower the minimal energy of the layer for the main
part of the wing, i.e. for $\omega_f\leq\omega_{bend}^{(j)}$ where
$\omega_{bend}^{(j)}$ is defined by the condition $\Delta
E_{r,NR}^{(j)}= \max(|\Delta E|)\equiv \sqrt{2}h$. For
$\omega_f>\omega_{bend}^{(j)}$, the minimal energy in the layer
occurs at $\tilde{\psi}=\pi$, and it is determined exclusively by
the variation of energy during the velocity pulse (the NR
contribution is close to zero at such $\tilde{\psi}$). Thus, we
conclude that there is a bending of the wing at
$\omega_f=\omega_{bend}^{(j)}$:

\begin{eqnarray}
&&\Delta E_r^{(j)}(\omega_f)= \Delta E_{r,NR}^{(j)}(\omega_f),
\quad\quad\omega_{\max}^{(j)}<\omega_f\leq\omega_{bend}^{(j)},\nonumber\\
&&\Delta E_r^{(j)}(\omega_f)= \sqrt{2}h,
\quad\quad\omega_f\geq\omega_{bend}^{(j)},\nonumber\\
&&\omega_{bend}^{(j)}=\frac{2\pi j}{\ln(8\sqrt{2}/h)+1-2\pi},
\end{eqnarray}

\noindent where $\Delta E_{r,NR}^{(j)}(\omega_f)$ is given by Eqs.
(54) and (55).

Analogously to the previous case, $\Delta
E^{(-)}(\omega_f)$ may be approximated in the whole frequency range by Eq. (41) with $\Delta E_l^{(j)}$ and $\Delta E_r^{(j)}$ given by Eqs. (58) and (60) respectively. Moreover, unlike the previous case, now the theory accurately describes also the range far beyond the peaks: $\Delta
E^{(-)}$ is dominated in this range by the velocity pulse contribution $\Delta
E$, which is accurately taken into account both by Eq. (58) and by Eq. (60).

Fig. 5 shows a very reasonable agreement between theory and
simulations, especially for the 1st peak \cite{2nd peak}.

\begin{figure}[tb]
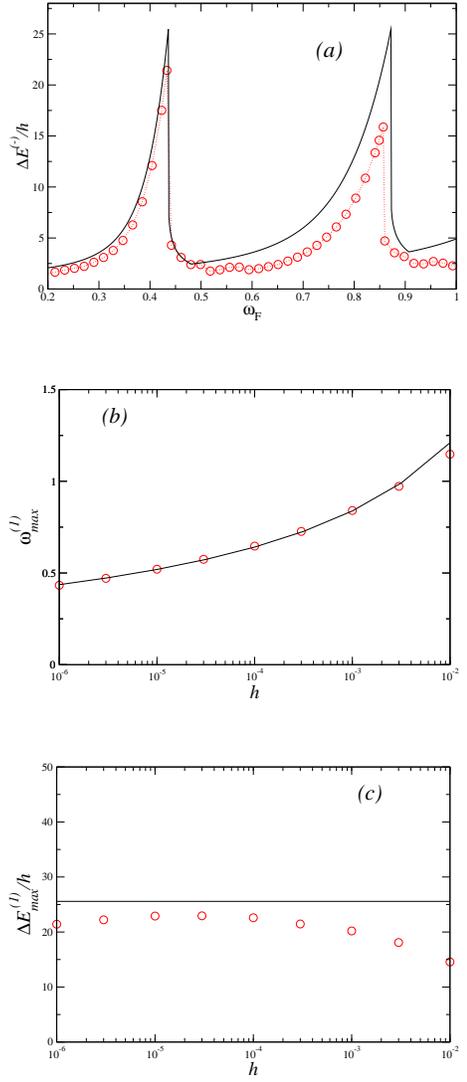

\includegraphics*[width = 6 cm]{Fig5a.eps}
\vskip 0.8 cm
\includegraphics*[width = 6 cm]{Fig5b.eps}
\vskip 0.8 cm
\includegraphics*[width = 6 cm]{Fig5c.eps}
\vskip 0.2 cm
 \caption{An archetypal example of type II: ac driven Duffing oscillator (42).
 Comparison of theory (solid lines) and simulations (circles):
 (a) the deviation $\Delta E^{(-)}(\omega_f)$ of the lower boundary of the chaotic
layer from the separatrix,
 normalized by the perturbation amplitude $h$, as a function of the perturbation frequency
 $\omega_f$, for $h=10^{-6}$; the theory is by Eqs. (41), (50), (51), (52), (54), (55), (58) and (60); (b) the frequency of the 1st maximum in $\Delta E^{(-)}(\omega_f)$
 as a function of $h$; the theory is by Eq. (50); (c) the 1st maximum in $\Delta E^{(-)}(\omega_f)/h$
 as a function of $h$; the theory is by Eq. (59).} \label{abd:fig5}
\end{figure}

\subsection{Pendulum with an oscillating suspension point}

Consider the archetypal Hamiltonian
\cite{abdullaev,gelfreich,treschev,shevchenko:1998,shevchenko}

\begin{eqnarray}
&& H=H_0+hV,
\nonumber\\
&&  H_0=\frac{p^2}{2}+\cos(q), \quad\quad
V=-\cos(q)\cos(\omega_ft),\nonumber\\
% \quad\quad
&& h\ll 1.
\end{eqnarray}

Though the treatment is similar to the previous case, there are
also characteristic differences. One of them is the following: although the
resonance Hamiltonian is similar to the Hamiltonian (16), instead
of the Fourier component of the coordinate, $q_n$, there should be
the Fourier component of $\cos(q)$, $V_n$, which can be shown to
be:

\begin{eqnarray}
&& V_{2j}\simeq  (-1)^{j+1}\frac{4}{\pi}\omega(E), \quad\quad
E_s-E\ll 1, \\
% \quad\quad
&& V_{2j-1}=0,
\nonumber\\
&& j=1,2,...\ll\frac{2\pi}{\omega(E)}, \nonumber\\&&
V_n\equiv\frac{1}{2\pi}\int_0^{2\pi}{\rm
d}\psi\cos(q)\cos(n\psi).\nonumber
\end{eqnarray}

The description of the chaotic layer of the separatrix map at the
lowest order, i.e. within the NR
approximation, is similar to that for the ac-driven Duffing
oscillator. So, we present only the results, marking them with the
subscript \lq\lq $NR$\rq\rq.

The frequency of the maximum of a given $j$-th peak is:

\begin{equation}
\omega_{\max}^{(j)}\simeq \frac{2\pi j}{\ln(4/h)},\quad\quad
j=1,2,...\ll\ln(4/h).
\end{equation}

\noindent This expression well agrees with simulations for the
Hamiltonian system (Fig. 6(b)). To logarithmic accuracy, Eq. (63)
coincides with the formula following from Eq. (8) of
\cite{shevchenko:1998} (reproduced in \cite{shevchenko} as Eq. (21))
taken in the asymptotic limit $h\rightarrow 0$ (or, equivalently,
$\omega_{\max}^{(j)}\rightarrow 0$). However, the numerical factor
in the argument of the logarithm in the asymptotic formula following
from the result of \cite{shevchenko:1998,shevchenko} is half our
value: this is because the
nonlinear resonance is approximated in \cite{shevchenko:1998,shevchenko} by the conventional pendulum
model which is not valid near the separatrix (cf. our Sec. III
above).

\begin{figure}[tb]
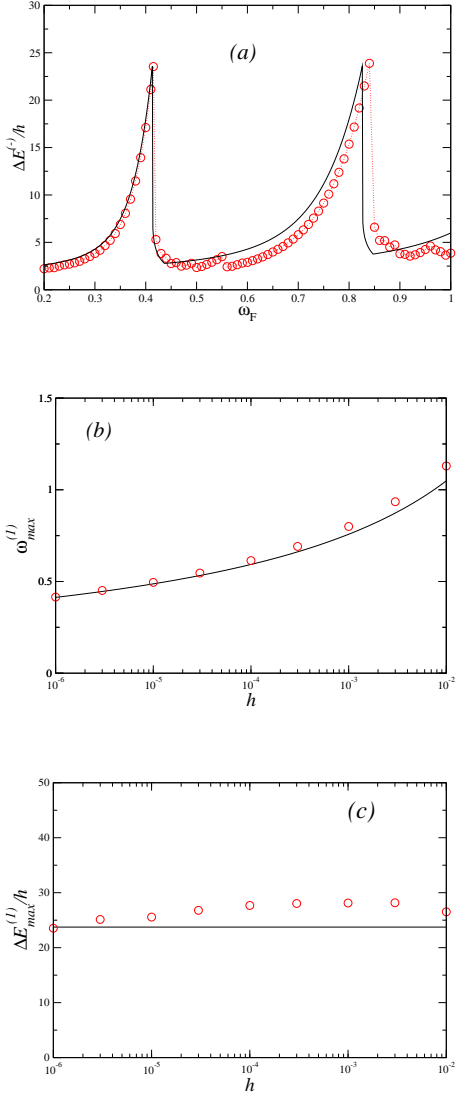

\includegraphics*[width = 6 cm]{Fig6a.eps}
\vskip 0.9 cm
\includegraphics*[width = 6 cm]{Fig6b.eps}
\vskip 0.9 cm
\includegraphics*[width = 6 cm]{Fig6c.eps}
\vskip 0.2 cm
 \caption{An archetypal example of type II: pendulum with an oscillating suspension
 point (61).
 Comparison of theory (solid lines) and simulations (circles):
 (a) the deviation $\Delta E^{(-)}(\omega_f)$ of the lower boundary of the chaotic
layer from the separatrix,
 normalized by the perturbation amplitude $h$, as a function of the perturbation frequency
 $\omega_f$, for $h=10^{-6}$; the theory is by Eqs. (41), (63), (64), (65), (67), (68), (71) and (73);
 (b) the frequency of the 1st maximum in $\Delta E^{(-)}(\omega_f)$
 as a function of $h$; the theory is by Eq. (63); (c) the 1st maximum in $\Delta E^{(-)}(\omega_f)/h$
 as a function of $h$; the theory is by Eq. (72).} \label{abe:fig6}
\end{figure}

The left wing of the $j$th peak of $\Delta E^{(-)}_{NR}(\omega_f)$
is described by the function

\begin{eqnarray}
&&\Delta E^{(j)}_{l,NR}(\omega_f)=32(1+y)\exp\left(-\frac{2\pi
j}{\omega_f}\right)
\\&&\quad\quad\quad\quad\quad\quad\equiv\frac{8
h}{\ln(1+y)-y/(1+y)},\quad\quad
\omega_f\leq\omega_{\max}^{(j)},\nonumber
\end{eqnarray}

\noindent where $y$ is the positive solution of the transcendental
equation

\begin{equation}
(1+y)\ln(1+y)-y=\frac{h}{4}\exp\left(\frac{2\pi
j}{\omega_f}\right),\quad\quad y>0.
\end{equation}

Similarly to the previous cases, $1+y(\omega_{\max}^{(j)})={\rm
e}$. Hence,

\begin{equation}
\Delta E^{(j)}_{\max,NR}={\rm e}
(E_s-E_r^{(j)}(\omega_{\max}^{(j)}))=8{\rm e} h.
\end{equation}

Eq. (66) confirms the rough estimate (13).

The right wing of the peak is described by the function

\begin{eqnarray}
&&\Delta E^{(j)}_{r,NR}(\omega_f)=32z\exp\left(-\frac{2\pi
j}{\omega_f}\right)\nonumber\\ &&\quad\quad \equiv\frac{8
h}{1+\ln(1/z)}, \quad\quad \omega_f>\omega_{\max}^{(j)},
\end{eqnarray}

\noindent where $z<1$ is the solution of the transcendental
equation

\begin{equation}
z(1+\ln(1/z))=\frac{h}{4}\exp\left(\frac{2\pi
j}{\omega_f}\right),\quad\quad 0<z<1.
\end{equation}

\noindent Similarly to the previous cases,
$z(\omega_f\rightarrow\omega_{\max}^{(j)})\rightarrow 1$.

Consider now the variation of energy during the velocity pulse.
Though the final result looks quite similar to the case with
a single saddle, its derivation has some characteristic
differences, and we present it in detail. Unlike the case with
a single saddle, the pulse may start close either to the left
turning point or to the right turning point, and the sign of the
velocity in such pulses is opposite
\cite{Zaslavsky:1991,pre2008}. As concerns the angle $\psi$ in the
pulse, it is close to $-\pi/2$ or $\pi/2$ respectively. So, let us
calculate the change of energy from the beginning of the pulse,
$t_k$, till a given instant $t$ within the pulse:

\begin{eqnarray}
&&\Delta E =-\int_{t_k}^{t}{\rm d}\tau\dot{q}h\partial V/\partial
q=h\int_{t_k}^{t}{\rm
d}\tau\dot{q}(-\sin(q)\cos(\omega_f\tau))\nonumber\\
&&\simeq  h\cos(\omega_ft_k)\int_{t_k}^{t}{\rm
d}\tau\dot{q}(-\sin(q))=h\cos(\omega_ft_k)\cos(q)\left|_{t_k}^{t}\right.\nonumber\\
&&\simeq  h\cos(\omega_ft_k)(\cos(q(t))-1) .
\end{eqnarray}

\noindent Here, the third equality  assumes adiabaticity while the
last equality takes into account that the turning points are close
to the maxima of the potential i.e. close to a multiple of $2\pi$
(where the cosine is equal to 1).

The quantity $\Delta E$ (69) has the maximal absolute value at
$q=\pi$. So, we shall further consider

\begin{eqnarray}
&&\Delta E_{\max} =-2h\cos(\omega_ft_k)\equiv
-2h\cos(2j\psi_k-\tilde{\psi}_k)\nonumber\\
&&=(-1)^{j+1} 2h\cos(\tilde{\psi}_k).
\end{eqnarray}

\noindent The last equality takes into account that, as mentioned
above, the relevant $\psi_k$ is either $-\pi/2$ or $\pi/2$.

For the left wing, the value of $\tilde{\psi}$ at which the
chaotic layer of the separatrix map possesses a minimal energy
corresponds to the minimum of the resonance separatrix. It is
equal to $\pi$ or $0$ if the Fourier coefficient $V_{2j}$ is
positive or negative, i.e. for odd or even $j$,
respectively: see Eq. (63). Thus $\Delta E_{\max}=-2h$ for any $j$
and, therefore, it does lower the minimal energy of the boundary.
We conclude that

\begin{equation}
\Delta E_l^{(j)}(\omega_f)\simeq  \Delta
E_{l,NR}^{(j)}(\omega_f)+2h,\quad\quad\omega_f\leq\omega_{\max}^{(j)},
\end{equation}

\noindent where $\Delta E_{l,NR}^{(j)}(\omega_f)$ is given by Eqs.
(64)-(65). In particular, the maximum of the peak is:

\begin{equation}
\Delta E^{(j)}_{\max}\simeq  (4{\rm e}+1)2h\approx 23.7h.
\end{equation}

The expression (72) confirms the rough estimate (13) and well agrees
with simulations (Fig. 6(c)). At the same time, it differs from the
formula which can be obtained from Eq. (10) of
\cite{shevchenko:1998} (using also Eqs. (1), (3), (8), (9) of
\cite{shevchenko:1998}) in the asymptotic limit $h\rightarrow 0$:
the latter gives for $\Delta E_{\max}^{(j)}$ the asymptotic value
$32h$. Though the result \cite{shevchenko:1998} (referred also in
\cite{shevchenko}) provides for the correct functional dependence on
$h$, it is quantitatively incorrect because (i) it is based on the
pendulum approximation of the nonlinear resonance while this
approximation is not valid in the vicinity of the separatrix (see the discussion of this issue in
Sec. III above), and (ii) it does not take into account the
variation of energy during the velocity pulse.

The right wing, analogously to the case of the Duffing oscillator,
possesses a bending at $\omega_f=\omega_{bend}^{(j)}$ at which
$\Delta E_{r,NR}^{(j)}= |\Delta E_{\max}|\equiv 2h$, that
corresponds to the switching of the relevant $\tilde{\psi}$ by $\pi$.
We conclude that:

\begin{eqnarray}
&&\Delta E_r^{(j)}(\omega_f)= \Delta E_{r,NR}^{(j)}(\omega_f),
\quad\quad\omega_{\max}^{(j)}<\omega_f\leq\omega_{bend}^{(j)},\nonumber\\
&&\Delta E_r^{(j)}(\omega_f)= 2h,
\quad\quad\omega_f\geq\omega_{bend}^{(j)},\nonumber\\
&&\omega_{bend}^{(j)}=\frac{2\pi j}{\ln(16/h)-3},
\end{eqnarray}

\noindent where $\Delta E_{r,NR}^{(j)}(\omega_f)$ is given by Eqs.
(66) and (67).

Similarly to the previous case, both the peaks and the frequency
ranges far beyond the peaks are well approximated by Eq. (41) with
$\Delta E_l^{(j)}$ and $\Delta E_r^{(j)}$ given by Eqs. (71) and
(73) respectively (Fig. 6(a)).

\section{Estimate of the next-order corrections}

We have explicitly calculated only the leading
term 
$\Delta E$
in the  asymptotic expansion of the chaotic layer width.
The explicit calculation of the next-order term 
$\Delta E^{(next)} $
is possible but it is rather
complicated and cumbersome: see the closely related case with two
separatrices \cite{pre2008}, where most of the next-order contributions are
calculated quantitatively \cite{second}. In the present paper,
where the
perturbation amplitude $h$ in the numerical examples is 4 orders of magnitude smaller
than that in \cite{pre2008}, there is no particular need to
calculate the next-order
%corrections
correction $C$
quantitatively.
Let us estimate it just qualitatively, with the main purpose to
demonstrate that its ratio to the lowest-order term does vanish in
the asymptotic limit $h\rightarrow 0$.

We shall consider separately the
%contribution
contribution $\Delta E^{(next)}_{w}$
stemming from the
various corrections
%within
{\it within}
the resonance approximation (16) and
the
%contribution
contribution $\Delta E^{(next)}_{t}$
stemming from the corrections
%to
{\it to}
the resonance
approximation.

The former
%correction
contribution
may be estimated similarly to
the case considered in \cite{pre2008}:
it stems from the deviation of the GSS curve from the separatrix
(this deviation reaches $\delta$ at certain angles: see Eq. (7)), from the difference
between the exact resonance condition (20) and the approximate one
(21), etc. It can be shown that the absolute value
%$R_r$
%$R_{w}$
of the ratio between $\Delta E^{(next)}_w$ and the leading
term is logarithmically small (cf. \cite{pre2008}):

\begin{equation}
%R_r
\frac{|\Delta E^{(next)}_{w}|}{\Delta E}
\sim\frac{1}{\ln(1/h)}.
\end{equation}

Let us turn to the analysis of the contribution $\Delta E^{(next)}_{t}$, i.e.\
the contribution
stemming from the corrections to the resonance Hamiltonian (16). It is
convenient to consider separately the cases of the left and right wings of the
peak.

As described in Secs. IV and V above, the left wing corresponds in the
leading-order approximation to formation of the boundary of the layer by the
{\it separatrix} of the resonance Hamiltonian (16). The resonance approximation
(16) neglects time-periodic terms while the frequencies of oscillation of these
terms greatly exceed the frequency of eigenoscillation of the resonance
Hamiltonian (16) around its relevant elliptic point i.e. the elliptic point
inside the area limited by the resonance separatrix. As is well known
\cite{gelfreich,lichtenberg_lieberman,treschev,zaslavsky:1998,zaslavsky:2005,Zaslavsky:1991},
fast-oscillating terms acting on a system with a separatrix give rise to the
onset of an {\it exponentially narrow} chaotic layer in place of the
separatrix. In the present context, this means that the correction to the
maximal action $\tilde{I}$ stemming from fast-oscillating corrections to the
resonance Hamiltonian, i.e. $\Delta E^{(next)}_{t}$, is {\it exponentially
small}, thus being negligible in comparison with the correction $\Delta
E^{(next)}_{w}$ (see (74)).

The right wing, described in Secs. IV and V above, corresponds in
leading-order approximation to the formation of the boundary of the layer by
the resonance trajectory {\it tangent} to the GSS curve. For the part of the
right wing exponentially close in frequency to the frequency of the maximum,
the tangent trajectory is close to the resonance separatrix, so that the
correction stemming from fast-oscillating terms is exponentially small,
similarly to the case of the left wing. As the frequency further deviates from
the frequency of the maximum, the tangent trajectory further deviates from the
resonance separatrix and the correction $\Delta E^{(next)}_{t}$ differs from
the exponentially small correction estimated above. It may be estimated in the
following way.

It follows from the second-order approximation of the averaging method
\cite{bogmit} that the fast-oscillating terms lead, in the second-order
approximation, to the onset of additional terms
$h^2T_{\tilde{I}}(\tilde{I},\tilde{\psi})$  and
$h^2T_{\tilde{\psi}}(\tilde{I},\tilde{\psi})$ in the dynamic equations for slow
variables $\tilde{I}$ and $\tilde{\psi}$ respectively, where
$T_{\tilde{I}}(\tilde{I},\tilde{\psi})$ and
$T_{\tilde{\psi}}(\tilde{I},\tilde{\psi})$ are of the order of the
power-law-like function of $1/\ln(1/h)$ in the relevant range of $\tilde{I}$.
The corresponding correction to the width of the chaotic layer in energy may be
expressed as

\begin{equation}
\Delta E^{(next)}_{t}=\int_{t_{\min}}^{t_{\max}}{\rm
d}t\;h^2T_{\tilde{I}}\omega(\tilde{I}),
\end{equation}

\noindent where $t_{\min}$ and $t_{\max}$ are instants corresponding to the
minimum and maximum deviation of the tangent trajectory from the separatrix of
the unperturbed system (cf. Figs. 1(c) and 4(c)). The interval
$t_{\max}-t_{\min}$ may be estimated as follows:

\begin{equation}
t_{\max}-t_{\min}\sim\frac{\pi}{|<\dot{\tilde{\psi}}>|},
\end{equation}

\noindent where $<\dot{\tilde{\psi}}>$ is the value of $\dot{\tilde{\psi}}$
averaged over the tangent trajectory. It follows from (16) that

\begin{equation}
|<\dot{\tilde{\psi}}>|\sim\omega_f-\omega(E_s-\delta)
\sim\frac{\omega(E_s-\delta)}{\ln(1/h)}\sim
\frac{\omega_0}{\ln^2(1/h)}.
\end{equation}

Taking together Eqs. (75)-(77) and allowing for the fact that $T_{\tilde{I}}$
is of the order of a power-law-like function of $1/\ln(1/h)$, we conclude that

\begin{equation}
\Delta E^{(next)}_{t}\sim h^2P(\ln(1/h)),
\end{equation}

\noindent where $P(x)$ is some power-law-like function.

The value $\Delta E^{(next)}_{t}$ is still asymptotically smaller than the
absolute value of the correction within the resonance approximation, $|\Delta
E^{(next)}_{w}|$, which is of the order of $h$ or $h/\ln(1/h)$ for systems of
type I or type II respectively.

Thus, we conclude that, both for the left and right wings of the peak, (i) the
correction $\Delta E^{(next)}$ is determined by the correction within the
resonance approximation $\Delta E^{(next)}_{w}$, and (ii) in the asymptotic
limit $h\rightarrow 0$, the overall next-order correction is negligible in
comparison with the leading term:

\begin{eqnarray}
&&\frac{|\Delta E^{(next)}|}{\Delta E}\equiv
\frac{|\Delta E^{(next)}_w+\Delta E^{(next)}_t|}{\Delta E}\approx
\frac{|\Delta E^{(next)}_w|}{\Delta E}
\sim
\nonumber
\\
&&
\quad\quad\sim\frac{1}{\ln(1/h)}
\stackrel{h\rightarrow
0}{\longrightarrow}0. 
\end{eqnarray}

\noindent This estimate well agrees with results in Figs.\ 3-6.

\section{Discussion}

In this section, we briefly discuss the following issues: 1) the {\it scaled} asymptotic shape of the
peaks, 2) peaks in the
range of {\it moderate} frequencies, 3) {\it steps} in the amplitude dependence
of the layer width, 4) an application to the {\it global chaos onset}.

\begin{itemize}

\item[1.]
Let us analyse the scaled asymptotic shape of the peaks. Consider first
systems of type I. Then the peaks are described in the
leading-order approximation exclusively within the separatrix map
dynamics (approximated, in turn, by the NR dynamics). As follows
from Eqs. (32), (34), (36), (39) and (40), most of the peak with a
given $j$ can be written in the universal scaled form:

\begin{equation}
\Delta E^{(j)}(\omega_f)=\Delta
E^{(j)}_{\max}S\left(\frac{\pi(2j-1)}{(\omega_{\max}^{(j)})^2}(\omega_f-\omega_{\max}^{(j)})
\right),
\end{equation}

\noindent where the universal function $S(\alpha)$ is strongly
asymmetric:

\begin{eqnarray}
&& S(\alpha)=\left\{^{S_l(\alpha)\quad {\rm for}\quad\alpha\leq 0,
}_{S_r(\alpha)\quad {\rm for}\quad\alpha > 0,}\right.
\\
&& S_l(\alpha)=\frac{1}{{\rm e}(\ln(1+y)-y/(1+y))},\nonumber\\
&&\quad\quad(1+y)\ln(1+y)-y=\exp(-\alpha),\nonumber\\
&& S_r(\alpha)=\frac{1}{{\rm e}(1+\ln(1/z))},\nonumber\\
&&\quad\quad z(1+\ln(1/z))=\exp(-\alpha).\nonumber
\end{eqnarray}

It is not difficult to show that

\begin{eqnarray}
&& S_l(\alpha=0)=1,\quad\quad\quad\quad\quad S_r(\alpha\rightarrow
+0)={\rm e}^{-1},
\\
&&\frac{{\rm d}S_l(\alpha= 0)}{{\rm d}\alpha}=1-{\rm
e}^{-1},\quad\frac{{\rm d}S_r(\alpha \rightarrow +0)}{{\rm
d}\alpha}\rightarrow -\infty,\nonumber
\\
&& S_l(\alpha\rightarrow -\infty)\propto \frac{1}{|\alpha|},\quad\quad\quad\quad\quad
S_r(\alpha\rightarrow \infty)\propto \frac{1}{\alpha}.\nonumber
\end{eqnarray}

Thus, the function $S(\alpha)$ is discontinuous at the maximum. To
the left of the maximum, the function relatively {\it slowly} approaches
the far part of the wing (which falls down power-like)
while, to the right of the maximum, the function first drops
{\it jump-wise} by a factor ${\rm e}$ and then {\it sharply} approaches
the far part of the wing (which falls down power-like).

As follows from Eqs. (80), (81), (82) and (27), the peaks are
logarithmically narrow, i.e. the ratio of the half-width of the
peak, $\Delta \omega^{(j)}$, to $\omega^{(j)}_{\max}$ is
logarithmically small:

\begin{equation}
\frac{\Delta
\omega^{(j)}}{\omega^{(j)}_{\max}}\sim\frac{1}{\ln\left(8(2j-1)/h\right)}.
\end{equation}

We emphasize that the shape (81) is not restricted to the example
(14): it is valid for any system of type I.

For systems of type II, the contributions from the NR and
from the
variation of energy during the pulse of velocity, as concerns the $h$
dependence, are formally of
the same order but, numerically,
the latter contribution is
typically much smaller than the former one. Thus, typically, the
function (81) well approximates the properly scaled shape of the
major part of the peak for systems of type II too.

\item[2.]

The quantitative theory presented in the paper relates only to the
peaks of {\it small} order $n$ i.e. in the range of logarithmically
small frequencies. At the same time, the magnitude of the peaks is
still significant up to the frequencies of the order of one. This
occurs because, for the motion close to the separatrix, the order
of magnitude of the Fourier coefficients remains the same up to
logarithmically large numbers $n$. The shape of the peaks remains
the same but their magnitude decreases typically (but, in some cases,
it may even increase in some range of frequencies). The
quantitative description of this decrease as well as the analysis of
more sophisticated cases requires a generalization of our theory,
that will be presented elsewhere.

\item[3.]
Apart from the frequency dependence of the layer width, our theory
is also relevant for the amplitude dependence: it describes the
jumps \cite {soskin2000} in the dependence of the width on $h$ and
the transition between the jumps and the linear dependence. The
values of $h$ at which the jumps occur, $h_{jump}^{(j)}$, are
determined by the same condition which determines
$\omega_{\max}^{(j)}$ in the frequency dependence of the width. The
formulas relevant to the left wings of the peaks in the frequency
dependence describe the ranges $h>h_{jump}^{(j)}$ while the formulas
relevant to the right wings describe the ranges $h<h_{jump}^{(j)}$.

\item[4.]
Finally we note that, apart from systems with a separatrix, our
work may be relevant to {\it nonlinear resonances} in any system. If the
system is perturbed by a weak time-periodic perturbation, then
nonlinear resonances arise and their dynamics is described by the
model of the auxiliary time-periodically perturbed pendulum
\cite{Chirikov:79,lichtenberg_lieberman,Zaslavsky:1991,zaslavsky:1998,zaslavsky:2005,abdullaev,gelfreich}.
If the original perturbation has a single harmonic, then the
effective perturbation of the auxiliary pendulum is necessarily a
high-frequency one, and chaotic layers associated with the
resonances are exponentially narrow
\cite{Chirikov:79,lichtenberg_lieberman,Zaslavsky:1991,zaslavsky:1998,zaslavsky:2005,abdullaev,gelfreich}
while our results are irrelevant. But, if either the amplitude or
the angle of the original perturbation is slowly modulated, or if
there is an additional harmonic of a slightly shifted frequency,
then the effective perturbation of the auxiliary pendulum is a
low-frequency one \cite{pre2008} and the layers become much wider
\cite{20} while our theoretical approach becomes relevant. It may
allow to find optimal parameters of the perturbation for the
facilitation of the onset of global chaos associated with the
overlap in energy between different-order nonlinear resonances
\cite{Chirikov:79}: the overlap may be expected to occur at a much
smaller amplitude of perturbation in comparison with that one
required for the overlap in case of a single-harmonic
perturbation.
%Details will be presented elsewhere.
\end{itemize}

\section{Conclusions}

We
have further developed the new \cite{pre2008} approach to the treatment of the separatrix chaos in the range of logarithmically small and moderate frequencies, where the chaos takes the largest possible area in phase space. The approach is based on the matching between the discrete chaotic dynamics of the SM and the continuous regular-like dynamics of the resonance Hamiltonian. Using this approach and taking also into account the dynamics in between instants corresponding to the SM, we
have presented the {\it first ever} accurate asymptotic description of high sharp peaks of the width of the separatrix chaotic
layer in energy as function of the
frequency of a weak time-periodic perturbation, including in particular the {\it absolute maximum} of the function.
Our work provides
the accurate base to explain former numerical and heuristic results and
intuitive assumptions
\cite{shevchenko:1998,shevchenko,vecheslavov,soskin2000,pre2008,proceedings},
corrects the errors of a previous heuristic theory
\cite{shevchenko:1998,shevchenko}, discovers new important
features, and opens up new horizons for future studies and applications.

The observed peaks arise due to the involvement of the nonlinear resonance
dynamics into the separatrix chaotic motion.
In the context of the magnitude of the peaks, all systems are classified into {\it two types}: the magnitude of the
peaks is proportional to the perturbation amplitude $h$ times
either a {\it logarithmically large} factor $\propto \ln(1/h)$ (for
systems of type I) or a {\it numerical} factor (for systems of type II).
Type I includes systems for which the separatrix of their
unperturbed Hamiltonian has more than one saddle while the perturbation
is not identical on adjacent saddles (an example is an ac-driven pendulum). All other systems belong to
type II. The latter type includes, in particular, a
pendulum with an oscillating suspension point, for which our
result differs from the result \cite{shevchenko:1998} since the latter (i) is
based on the conventional approximation of the nonlinear resonance
(not valid near the separatrix), and (ii) does not take into account the variation
of energy during the velocity pulse (i.e. in between the instants relevant to the separatrix map). Our
theory is verified by computer simulations.

The shape of the peaks is strongly asymmetric. In the asymptotic
limit of small amplitudes, the shape of the peaks for type I is
{\it universal}. For type II, the shape is quite similar, differing only
by a typically small contribution stemming from the variation of
energy during the velocity pulse.

Our theory describes the {\it jumps} of the width as a function of
the perturbation amplitude $h$ as well as the transition between the
jumps and the linear dependence.

Finally, our work suggests a new method for the {\it facilitation of global
chaos onset}
due to the enhanced overlap
of nonlinear resonances. The theoretical approach developed by us may be used to
derive the optimal choice of parameters of the perturbation leading
to the facilitation.

\begin{acknowledgments}
This work was partly supported by the grant within the Convention
between the Institute of Semiconductor Physics and University of
Pisa for 2008 and by the Royal Society grant 2007/R2-IJP. We acknowledge
discussions with Vassili Gelfreich, Igor Khovanov and Oleg
Yevtushenko. We especially appreciate numerous stimulating
discussions with George Zaslavsky and his role for the subject of
Hamiltonian chaos on the whole. We dedicate this paper to the
memory of George Zaslavsky who recently passed away.
\end{acknowledgments}


\begin{thebibliography}{00}
\bibitem{22_prime} Formally, ${\rm sgn}(E-E_s)$ is not defined for $E=E_s$ but, if to shift $E$ from $E_s$ for an infinitesemal value, ${\rm sgn}(E-E_s)$ acquires the value equal to either $+1$ or $-1$, dependently on the sign of the shift. 
Given that $\sigma_{i+1}$ is proportional to ${\rm sgn}(E_s-E_{i+1})$ while the latter is random-like (as it has been shown above),
$\sigma_{i+1}$ is not correlated with $\sigma_{i}$ if $E_{i+1}=E_s\pm 0$.
\bibitem{24_prime} For the SM relating to the ac-driven spatially periodic systems,
the time during which the SM undergoes a regular-like evolution above the separatrix diverges
in the adiabatic limit $\omega_f\rightarrow 0$ \cite{13_prime}, and
the width of the part of the SM layer above the separatrix diverges too.
However, we do not consider this case here since it is irrelevant to the main subject of the present paper i.e. to to the involvement of the resonance dynamics into the separatrix chaotic motion.
\bibitem{22_prime_prime} The GSS curve corresponds to the step of the SM which follows the state with $E=E_s$, as described above.
\bibitem{footnote_new} Namely, that of these two intervals is relevant in which the derivative ${\rm d}E/{\rm d}t$ in the nonlinear resonance equations (see Eq. (16) below) is positive or negative, for the case of the upper or lower GSS curve respectively.
\bibitem {29_prime_prime} The adiabatic estimate for $\Delta E^{(-)}(\omega_f)$ is equal approximately to $2\pi$. The estimate following from the results of the mathematical work \cite{treschev} for $\omega_f\sim 1$ is of the same order, so that it is conditionally represented in the inset in Fig. 2 by the same line as for the adiabatic estimate (dashed line).
\bibitem {27_prime} The only essential difference consists in that $q_n$ at the saddle is described by Eq. (46) rather than by Eq. (18).
\bibitem {2nd peak} The disagrement between the theory and
simulations for the magnitude of the 2nd peak is about three times
larger than that for the 1st peak, so that the height of the 2nd
peak is about 30$\%$ smaller than that by the asymptotic theory.
This occurs because, for energies relevant to the 2nd peak, the
deviation from the separatrix is much higher than that for the 1st
peak. Due to the latter, the Fourier coefficient $q_2(E)$ for the
relevant $E$ is significantly smaller than that by the asymptotic
formula (42). Besides, the velocity pulse contribution $\Delta E$ also
significantly decreases while the separatrix split increases as $\omega_f$
becomes $\sim 1$. The generalization of the theory for $\omega_f\sim 1$ will
be presented elsewhere.
\end{thebibliography}
\end{document}